\documentclass[10pt,a4paper]{article}
\usepackage[T1]{fontenc}
\usepackage{lmodern}
\usepackage[english]{babel}
\usepackage[hmargin=2.2cm,vmargin=2.5cm]{geometry}
\usepackage{amsmath}
\usepackage{amssymb}
\usepackage[nopatch]{microtype}
\usepackage{graphicx}
\usepackage{booktabs}
\usepackage{array}
\usepackage{float}
\usepackage{enumitem}
\usepackage[online]{threeparttable}
\usepackage[font=small,labelfont=bf,labelsep=period]{caption}
\usepackage{xurl}
\usepackage{tikz}
\usetikzlibrary{positioning,arrows.meta}
\usepackage[section]{placeins}
\usepackage{csquotes}
\usepackage[style=apa,backend=biber,natbib]{biblatex}
\usepackage[hypertexnames=false,hidelinks,pdfauthor={Colin Priest},pdftitle={Portfolio-aware scenario transfer of reserve movements: evidence from Lloyd's syndicates}]{hyperref}

\DeclareFieldFormat[article]{volume}{\mkbibbold{\apanum{#1}}}
\DeclareFieldFormat[periodical]{volume}{\mkbibbold{\apanum{#1}}}

\newcommand{\headrow}{}
\AtBeginEnvironment{table}{\footnotesize}
\AtBeginEnvironment{tablenotes}{\scriptsize}
\newcommand{\supp}[1]{Supplement~S#1}
\makeatletter
\def\@seccntformat#1{\@ifundefined{#1@cntformat}{\csname the#1\endcsname\quad}{\csname #1@cntformat\endcsname}}
\def\section@cntformat{\thesection.\quad}
\renewcommand\appendix{\par
  \setcounter{section}{0}
  \setcounter{subsection}{0}
  \gdef\thesection{Appendix \@arabic\c@section}}
\makeatother

\author{Colin Priest\\[3pt]
  {\small School of Risk and Actuarial Studies, University of New South Wales,}\\
  {\small Sydney, NSW 2052, Australia}\\
  {\small ORCID: 0009-0006-6786-3857. Email: c.priest@unsw.edu.au}}
\date{}

\title{Portfolio-aware scenario transfer of reserve movements: evidence from Lloyd's syndicates}
\newcommand{\pnum}{\ignorespaces}
\newcommand{\thekeywords}{Diversification; Lloyd's syndicates; Prior-year development; Reserve risk; Scenario transfer}
\AtEndEnvironment{abstract}{\par\medskip\noindent\textbf{Keywords:} \thekeywords}
\AtBeginDocument{\maketitle}

\begin{document}

\def\slashpathii#1/#2\relax{#1\ifx\\#2\\\else/\allowbreak\slashpathii#2\relax\fi}
\newcommand{\slashpath}[1]{\texttt{\expandafter\slashpathii\detokenize{#1}/\relax}}

\begin{abstract}
A claims-development movement on one portfolio may not transfer directly as a stress to another size. We estimate how to transfer such scenarios between
portfolio sizes, using 632 syndicate-years from 117 Lloyd's syndicates over 2015--2024. The measured movement is
the change in gross claims estimates on selected older underwriting cohorts, divided by opening gross claims
outstanding plus reserves taken on in the year whose development it includes; it is not a pure measure of
opening-reserve deterioration. Its dispersion falls materially with
reserve size, and a size-based scale predicts held-out syndicates far better than a single market scale. But the
transferred upper tail understates the smallest books: in the smallest reserve-size quartile, six of 158 held-out
observations exceeded the transferred 99.5th percentile, against 0.79 expected,
from three syndicates. Square-root scaling has descriptively lower held-out pinball loss at 99\% and 99.5\%, though more misses among the
largest books. For a \pounds500m
portfolio the transferred empirical 99.5th percentile is 0.34 of reserves ($95\%$ credible interval, conditional on
the fitted model, 0.23--0.52), against 0.67 in the raw market pool. Both are Syndicate~5820's 2016 movement. The contribution is inspectable scenario comparability, not a
calibrated capital quantile or a next-year forecast.
\end{abstract}

\section{Introduction}
\label{sec:intro}
\pnum Suppose an insurer with \pounds500m of gross claims reserves wants to use the market's worst reserve
movements as stress scenarios. One candidate is Syndicate~5820's 2016 year. During that year the estimated
gross claims on its older underwriting years rose by an amount equal to 0.67 of the \pounds57m of claims
reserves it held at the start.
Should the \pounds500m book stress itself by 0.67 of its reserves? Across the market, small books' movements are
more dispersed than large books'. The transfer of this paper assumes that a movement is comparable between books once
it is standardised by each book's fitted scale. Under that assumption and the size law fitted here, 0.67 on the
\pounds57m book corresponds to about 0.34 on the \pounds500m one. That is the model's transfer, not an observed effect
of the same event on a larger book, and not a demonstrated capital requirement. So a movement observed on one
portfolio needs re-expressing on the target's basis, under a stated assumption, before it is compared with the
target's own assumptions or reused as a scenario. This paper estimates one way to do that, and tests where it works.

\pnum The question matters for capital. Under Solvency~II, retained in the United Kingdom as Solvency~UK
\autocite{PRAPS0224}, the solvency capital requirement is calibrated to the 99.5\% one-year value-at-risk (VaR) of
basic own funds \autocite{SolvencyII2009}. For reserves, one-year development compares the opening best estimate
with claims paid during the year plus the closing best estimate of outstanding claims; it measures change over one year,
not uncertainty over the full run-off \autocite{Merz2008,SzatkowskiDelong2021,MunroeZehnwirthGoldenberg2018}. A firm's own history is short in the
adverse tail, so calibration leans on pooled market data and tail extrapolation \autocite{CEIOPS2010,IAIS2024}.
In its guidance for new syndicates, Lloyd's Standard Model uses net volatilities calibrated to an average market
portfolio; Lloyd's notes that these may not suit an unusual book \autocite{LloydsNewSyndicate2024}.

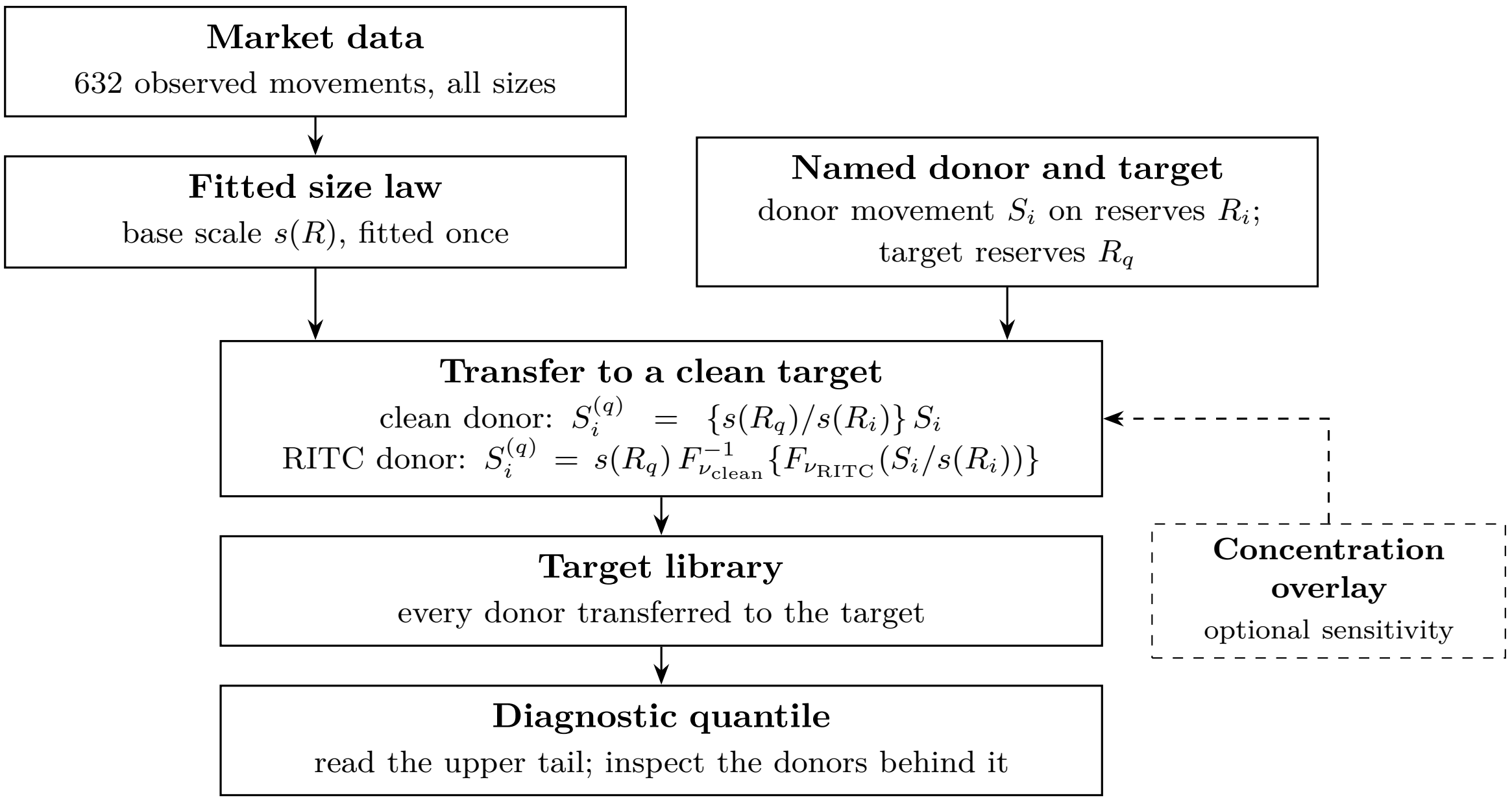
\begin{figure}[H]
\centering
\begin{tikzpicture}[
  wfbox/.style={draw, semithick, align=center, text width=0.36\linewidth, inner sep=5pt},
  midbox/.style={draw, semithick, align=center, text width=0.52\linewidth, inner sep=5pt},
  optbox/.style={draw, dashed, align=center, text width=0.20\linewidth, inner sep=4pt},
  wfarrow/.style={-{Stealth[length=6pt]}, semithick},
  node distance=11pt and 20pt
]
\node[wfbox] (data) {\textbf{Market data}\\[1pt]
  {\small 632 observed movements, all sizes}};
\node[wfbox, below=of data] (law) {\textbf{Fitted size law}\\[1pt]
  {\small base scale $s(R)$, fitted once}};
\node[wfbox, right=of law] (target) {\textbf{Named donor and target}\\[1pt]
  {\small donor movement $S_i$ on reserves $R_i$;\\ target reserves $R_q$}};
\path (law.south) -- (target.south) coordinate[pos=0.5] (mid);
\node[midbox, below=18pt of mid] (transfer) {\textbf{Transfer to a clean target}\\[1pt]
  {\small clean donor: $S^{(q)}_i=\{s(R_q)/s(R_i)\}\,S_i$\\
  RITC donor: $S^{(q)}_i=s(R_q)\,F^{-1}_{\nu_{\mathrm{clean}}}\{F_{\nu_{\mathrm{RITC}}}(S_i/s(R_i))\}$}};
\node[midbox, below=of transfer] (library) {\textbf{Target library}\\[1pt]
  {\small every donor transferred to the target}};
\node[midbox, below=of library] (diag) {\textbf{Diagnostic quantile}\\[1pt]
  {\small read the upper tail; inspect the donors behind it}};
\node[optbox, right=14pt of library] (overlay) {\small\textbf{Concentration overlay}\\
  {\footnotesize optional sensitivity}};
\draw[wfarrow] (data) -- (law);
\draw[wfarrow] (law.south) -- (transfer.north -| law.south);
\draw[wfarrow] (target.south) -- (transfer.north -| target.south);
\draw[wfarrow] (transfer) -- (library);
\draw[wfarrow] (library) -- (diag);
\draw[wfarrow, dashed] (overlay.north) |- (transfer.east);
\end{tikzpicture}
\caption{The scenario-transfer workflow. The size law is fitted once, on the whole market sample.
Each named donor movement is then transferred to a target in the default clean regime. A clean donor's movement is
multiplied by the ratio of fitted base scales. A donor in the reinsurance-to-close (RITC) regime is first divided by
its base scale and mapped to the clean regime's quantile at the same Student-$t$ rank, and only then multiplied by the
target's base scale (Section~\ref{sec:framework:operator}). For an RITC-affected target the roles reverse: clean donors
are mapped and RITC donors transfer by the ratio.
Repeating this for every donor gives the target library, whose upper quantile is read as a diagnostic and whose
supporting donors are inspected. The concentration overlay is an optional sensitivity
(Section~\ref{sec:framework:concentration}).}
\label{fig:workflow}
\end{figure}

\pnum The nearest antecedent is volatility-adjusted historical simulation \autocite{HullWhite1998}, later
filtered historical simulation \autocite{BaroneAdesiGiannopoulosVosper1999}, which replays a past market return
at today's volatility, $r^*_t=(\sigma_{\mathrm{now}}/\sigma_t)\,r_t$. Our transfer of a clean donor, one outside the
reinsurance-to-close (RITC) regime, has the same form, $S^{(q)}_i=\{s(R_q)/s(R_i)\}\,S_i$, but it reuses a standardised shock across portfolios of different size,
not across time, and the scale $s(\cdot)$ is a fitted function of reserve size. Related work addresses neighbouring
questions.
Stochastic reserving estimates a portfolio's own reserve distribution \autocite{AvanziTaylorVu2016}; credibility
theory pools parameters across risks, not realised movements \autocite{FreesWang2006}; and operational-risk
practice scales external losses by firm size without a fitted law for reserve movements
\autocite{ShevchenkoPeters2013}.

\pnum This study estimates that law from the public accounts of 632 syndicate-years from 117 Lloyd's syndicates,
transfers each historical movement donor by donor using the resulting scales and, where regimes differ, the tail map,
and tests the transfer on syndicates
held out of the fit. The measured movement is the development of selected older underwriting cohorts, which is not a
pure measure of deterioration in opening reserves. Size improves density prediction materially. But the transferred upper tail understates the
smallest books, and it has no demonstrated tail-score advantage over plain square-root scaling. So the method
supports inspectable scenario comparison; it does not establish a calibrated capital quantile.

\pnum The scope is narrow. This is not a reserving method: we fit no chain-ladder or other stochastic
reserving model and project no ultimates \autocite{Mack1993,EnglandVerrall2002}. Nor is it a forecast. The
validation holds out whole syndicates, so it asks how well the rest of the market describes a syndicate it has
not seen; it does not test a firm's next year with only the information available at the time. The same
transfer can also re-express a syndicate's own history on its current, changed book (\supp{24}).

\pnum Figure~\ref{fig:workflow} shows the workflow. Section~\ref{sec:data} defines the measured movement and the
sample. Section~\ref{sec:framework} builds the size law and the transfer. Section~\ref{sec:results} gives the
evidence, including where the transfer fails. Section~\ref{sec:vignettes} works one target through, and
Section~\ref{sec:implications} says what a practitioner can and cannot conclude.

\section{Data and the measured movement}
\label{sec:data}
Each observation is one syndicate's one-year movement in the estimated claims of its older underwriting cohorts,
scaled by its reserve base: the claims reserves it held at the start of the year, with any it took on during the year
whose development is measured. This section defines that quantity, the sample and the checks on its
extraction.

\subsection{Source, unit and size}
\label{sec:data:source}
\pnum The data are the publicly filed Lloyd's syndicate reports and accounts for reporting years 2014--2024.
Observations begin in 2015. The claims-development disclosure the measure needs dates from the move to Financial
Reporting Standards (FRS)~102 and~103, which Lloyd's required from the 2015 accounts with 2014 comparatives
\autocite{FRS103,LloydsUKGAAP2014}, and no 2014 record enters the working sample (\supp{7}).

\pnum The unit of analysis is a \emph{syndicate-year} $(i,t)$, one record for syndicate $i$ in reporting year $t$.
The size measure $R_{i,t}$ is the \emph{reserve base}: the opening gross claims outstanding, the unpaid claims
provision at the start of the year read from the balance-sheet claims-outstanding line, plus the gross claims
reserves the syndicate took on during the year when the year's development covers them. The take-on augments the
model's base; it is not part of the accounting opening balance. It applies to 8 of the 632 records. For example,
3500/2019 opened with 259.8m and took on 552.8m whose development its triangle includes, so its reserve base is
812.5m (\supp{9}); the three amounts are rounded independently. For the other 624 records the reserve base is the opening balance. The numerator $M_{i,t}$, the year's change in
the older cohorts' estimated claims, is defined in Section~\ref{sec:data:pyd}.

\pnum The ratio $M_{i,t}/R_{i,t}$ does not depend on the reporting currency, but $R$ does, so every amount is
converted to sterling at the reporting-date rate before the size law is fitted. About a quarter of the corpus
reports in US dollars. Reconverting all 162 dollar-reported openings at the prior year-end rate moves them by
$5.4\%$ on average and the worked stress of Section~\ref{sec:vignettes} by $0.1\%$ (\supp{11}). Amounts are quoted
in millions (m) or billions (bn) of pounds sterling (GBP).

\subsection{The measured movement}
\label{sec:data:pyd}
\pnum For a syndicate-year $(i,t)$ with reserve base $R_{i,t}$, the \emph{measured movement}
(or severity) is
\begin{equation}\label{eq:severity}
S_{i,t} = \frac{M_{i,t}}{R_{i,t}},
\qquad
M_{i,t} = \sum_{u\in\mathcal U_{i,t}}\Bigl(\hat U^{(t)}_{i,u}-\hat U^{(t-1)}_{i,u}\Bigr),
\end{equation}
where $\hat U^{(t)}_{i,u}$ and $\hat U^{(t-1)}_{i,u}$ are the estimates of underwriting year $u$'s ultimate gross
claims (paid plus outstanding) at the ends of years $t$ and $t-1$, ordinarily as the gross claims-development table in the
year-$t$ report prints them. Source-evidenced corrections use reconciled cells or bases where the printed table is
misstated; 3330/2017 instead uses a roll-forward estimate whose cohort scope remains uncertain (\supp{2}). The set $\mathcal U_{i,t}$ holds the underwriting years $u\le t-2$ for which that table
prints both estimates. A cohort the table does not develop, such as an older-years bucket, is outside
$\mathcal U_{i,t}$: it is not counted as a zero movement, but its reserves stay in $R_{i,t}$.

\pnum In words, $M$ is the change during year $t$ in the estimated ultimate claims of the older underwriting years
(the mature cohorts, $u\le t-2$),
and $S$ expresses it as a share of the reserve base: all the claims reserves held at the start of the year, with
any taken on during it whose development $M$ includes. Payments cancel in $M$: a
payment raises paid claims and lowers outstanding claims by the same amount. Positive $S$ means an increase in
estimated ultimate claims, and negative $S$ a decrease. Neither sign alone identifies a strengthening or release
of opening reserves. For example, Syndicate~5820's 2016 report prints gross
increases of 1.9m, 3.2m, 7.7m and 25.6m in the estimated ultimates of its 2011 to 2014 underwriting years, 38.4m in
all; over opening reserves of 57.1m this gives $S=0.67$ (\supp{1}). Table~\ref{tab:outcome} sets the quantity beside
three it is not.

\pnum $S$ measures the deterioration of opening claims reserves only when four conditions hold, and the public
accounts do not separate the parts when one fails:
\begin{enumerate}
  \item \emph{basis}: the cells are incurred or ultimate claims, not paid; a paid table would add the year's
        payments to $M$ and push $S$ up;
  \item \emph{earning}: the cohorts in $\mathcal U_{i,t}$ had finished earning by the opening date; if not, claims
        emerging on exposure still earning sit inside $M$ and push $S$ up;
  \item \emph{coverage}: the cohorts in $\mathcal U_{i,t}$ hold nearly all of $R$; reserves of the youngest cohort, or
        of a bucket the table does not develop, sit in $R$ but their movement is outside $M$. Including those reserves
        dilutes $|S|$ relative to dividing the same $M$ by the selected cohorts' reserves; the omitted movement could
        change the full-reserve deterioration in either direction;
  \item \emph{comparability}: $M$ and $R$ are on one gross basis and one currency translation; a reinsurers' share
        read as $R$ pushes $S$ up, and a table not retranslated between reports moves $S$ either way.
\end{enumerate}
So $S$ is observed cohort development. It can differ from pure deterioration of opening reserves through earning
and exposure emergence, cohort coverage and booked margins, most for books with prolonged premium earning, delayed
inception or continuing exposure in old underwriting years. Section~\ref{sec:results:smallbook} shows where this
matters, and \supp{1} and \supp{4} give the diagnostics for each condition.

\pnum Three construction rules complete the definition. The movements use booked provisions, so stated
management margins above the actuarial best estimate build and release inside them (\supp{10}). Where a syndicate
took business on in the report year and the development covers it, $R_{i,t}$ includes the gross claims reserves the
filing states were taken on, as Section~\ref{sec:data:source} defines the reserve base (\supp{9}). And a printed triangle figure is never replaced because a broader
movement has the other sign; an earlier 672-record construction did this, and its counts and refit are in \supp{5}.

\begin{table}[!t]
\begin{threeparttable}
\caption{The measured movement $S=M/R$ of Equation~\eqref{eq:severity}: what is rescaled, and three quantities it
is not.}
\label{tab:outcome}
\small
\begin{tabular}{p{0.27\linewidth}p{0.65\linewidth}}
\toprule
Numerator $M$ & Change during year $t$ in the estimated ultimate (paid plus outstanding) of the underwriting years
$u\le t-2$ the gross claims-development table develops, ordinarily read from that table with source-evidenced corrections. \\
Denominator $R$ & Reserve base: total opening gross claims outstanding at 31 December $t-1$, all underwriting
years, plus gross claims reserves taken on in year $t$ whose development $M$ includes (8 of the 632 records). \\
Basis and sign & Gross of reinsurance; booked provisions, including stated margins. Adverse is positive. \\
\midrule
Not: re-estimation of earned exposure & The change in claims estimates for exposure already earned at the opening
date, which is what the capital regime stresses. $M$ also holds emergence on exposure still earning in the selected
cohorts. \\
Not: calendar prior-year development & Calendar prior-year provision development. That figure includes the year $t-1$
cohort and is not partitioned by cohort. \\
Not: deterioration of the cohorts' own reserves & Deterioration as a share of the selected cohorts' own reserves. The
denominator here is the whole reserve base, including qualifying take-ons. \\
\bottomrule
\end{tabular}
\end{threeparttable}
\end{table}
\FloatBarrier

\subsection{Sample, source rule and assurance}
\label{sec:data:filters}
\pnum The sample is built in two steps. The first starts from the 902 corpus records that have a usable development
reading and opening reserves above \pounds0.1m and are not in run-off.\footnote{The corpus is not a stage of one
nested flow. Over 2014--2024 the market had 1{,}045 active syndicate-years, and automated collection retrieved
1{,}065 source reports (970 PDFs and 95 Hypertext Markup Language (HTML) filings); the two populations are reconciled in \ref{app:audit:exclusions} and \supp{7}.} It keeps those
with a gross-basis movement as Equation~\eqref{eq:severity} defines it, a computable $S$ and complete
line-of-business weights. For example, the gross-basis step sets aside 98 corpus records (82 recorded as net and 16 with
an unstated basis), and a premium mix is admitted only when its classes add to an independently read premium total
(Table~\ref{tab:reconcile}, \supp{8}). This gives 643 records from 118 syndicates. The second step applies the
source-history rule below, which sets 11 of them aside. The \emph{primary sample} is the remaining 632 records from
117 syndicates over the reporting years 2015--2024, and the 643-record sample is kept as a comparator.

\pnum Selection into the sample runs strongly by size. The included target records have a median reserve base of
\pounds421.5m, against \pounds31.2m for those left out (\ref{app:audit:exclusions}). So small syndicates are
under-represented, the fit leans on established books, and missing-at-random cannot be established for the small
books left out (\supp{6}).

\pnum The source-history rule sets an observation aside when specific source evidence raises a material concern,
left open by the filing, about overlapping claims history, gross/net mixing or cohort reassignment through closure,
and the declared gross original-cohort movement cannot be reconstructed or validated. It applies whatever the
movement's size or sign. Unexpired exposure, an ordinary internal reinsurance to close, a funds-withheld arrangement,
unusual business, youth or special-purpose status is not by itself a reason. So 6131/2020 and all six years of
Syndicate~1991 stay: their profiles limit where the measured movement applies (Section~\ref{sec:results:smallbook}),
but they do not breach the rule. The rule is post hoc. It was written down, before the census refits, after one record,
6131/2021, was found while investigating the small-book tail, and applied to the 101 records whose source features it
selects, chosen without reference to their movements and each read twice. It set aside 11 records; the same readings
corrected two figures and reversed one earlier correction, and the 643-record comparator carries all three
(\supp{2}). The census cannot find a concern a filing does not show.

\pnum Extraction is partly machine-assisted. The movement is ordinarily computed by arithmetic from the gross claims table the
filing prints in absolute amounts, with the source-evidenced corrections just described. The design uses two independent large language models to read report text, but not every report has two usable
model outputs. Where both supply a comparable field, their readings are compared. A disagreement is logged but excludes nothing (\ref{app:audit}). On 13 October 2026, before the
primary outcome was defined, a random draw of 100 records from the 400 of the then 674-record working sample that no
earlier reading had covered was read against their filings. Of the 100, 98 were adjudicable. It put the probability that a record's adopted development
figure was materially wrong at $2.5\%$ (posterior mean; $95\%$ credible interval $0.4\%$ to $6.4\%$;
\ref{app:audit:errorrate}), and its two errors were then repaired. That is a historical checkpoint of the extraction,
not a measured rate for the 632 records; the paper assumes it for records classified as unread by the simulation
of errors not yet found. It measures fidelity to the paper's numerator: a triangle figure checked against the triangle. It is not independent actuarial judgement, and it does not show that
the triangle measures the economic quantity of Section~\ref{sec:data:pyd}.

\pnum External reinsurance to close (RITC) and confirmed inward transfers of another syndicate's older liabilities
change the book whose development is measured and can introduce a transaction step. The RITC scan and confirmed-transfer review flag such cases (\supp{9}). The 27
syndicate-years in the sample that carry either (20 flagged by the RITC scan, 5 by a confirmed transfer alone and 2 by
a confirmed take-on alone) are kept. They are the \emph{RITC regime}: they have their own tail index in the fit, a
modelling choice the data do not resolve (Section~\ref{sec:results:posterior}), and a rank map in the transfer
(Sections~\ref{sec:framework:model} and~\ref{sec:framework:operator}). A \emph{clean} record
is one outside that regime; the word does not mean the record is free of measurement concerns.

\subsection{Business mix and its proxy}
\label{sec:data:lob}
\pnum Business mix enters only the concentration overlay of Section~\ref{sec:framework:concentration}. Syndicates do
not disclose reserves by line, so we use the disclosed premium shares $w_{i,t,\ell}$ as a proxy and summarise them by
the Herfindahl--Hirschman index (HHI) \autocite{Hirschman1945,Herfindahl1950}
\begin{equation}\label{eq:hhi}
H_{i,t} = \sum_{\ell} w_{i,t,\ell}^{2},
\end{equation}
with $1/H_{i,t}$ the effective number of equally weighted lines. $H$ describes current premium, not the business that
generated an old cohort's reserves, and it says nothing about dependence between lines. It is also coarse: the
harmonised taxonomy leaves the largest premium share in an unclassified \emph{Aggregate} bucket for 302 of the 632
records, $48\%$ (\supp{8}), and \supp{19} tests the overlay against perturbations of the proxy.

\section{The size law and scenario transfer}
\label{sec:framework}
A movement is transferred by rescaling it. This section explains the transfer with one real donor, then builds the
size law it uses, the fitted model that estimates that law, the full operator and the optional concentration overlay.

\subsection{Transfer by a ratio of scales}
\label{sec:framework:transfer}
\pnum Suppose the movements of a book with reserves $R$ are shocks of a common shape multiplied by a scale $s(R)$ that
depends on size. Then a donor movement $S_i$ observed on reserves $R_i$ is re-expressed on a target with reserves $R_q$
by
\begin{equation}\label{eq:transfer}
S^{(q)}_i = c_i\,S_i,
\qquad
c_i=\frac{s(R_q)}{s(R_i)}.
\end{equation}
The transferred movement keeps the donor's standardised shock $S_i/s(R_i)$ and gives it the target's scale. A donor
smaller than the target has the larger scale, so its movement shrinks; a larger donor's movement grows.

\pnum Take one real donor. Syndicate~5820's 2015 movement was $S_i=+0.155$ (a 15.5\% adverse movement) on opening
reserves of 67.0m. Under the size law fitted in Section~\ref{sec:framework:model}, its base scale is $s(67)=0.097$ and
a 500m target's is $s(500)=0.052$. So the transfer factor is $c_i=0.0523/0.0971=0.538$ (95\% credible interval 0.478
to 0.627), and the transferred movement is $0.538\times0.155=0.083$ (0.074 to 0.097). The 15.5\% stress on a 67m book
corresponds to an 8.3\% movement on a 500m target. The donor is clean, so nothing else changes.

\pnum Equation~\eqref{eq:transfer} has the form of volatility-adjusted historical simulation, applied across
portfolios. Two things remain to be supplied: an estimate of $s(R)$, and a rule for the 27 donors in the RITC regime,
whose tails may differ. The other 605 donors (about 96\% of the pool) are clean, and for them
Equation~\eqref{eq:transfer} is the whole operator.

\subsection{Pooling benchmarks and the size law}
\label{sec:framework:size}
\pnum Classical risk pooling supplies two benchmarks for the scale of the aggregate movement $M$. If $M$ were the sum
of many independent, similarly sized blocks with finite variance, its standard deviation would grow as $R^{1/2}$,
the $\sqrt N$ result. Comonotonic (fully dependent) blocks give growth proportional to $R$, with no diversification
\autocite{Dhaene2002a,Dhaene2002b}. Writing the scale of $M$ as proportional to $R^k$, the scale of $S=M/R$ is
proportional to $R^{k-1}$: $k=\tfrac12$ gives square-root scaling of $S$, and $k=1$ a scale that does not depend on
size.\footnote{The same power-law scaling is what reserving variance functions encode (for instance the Tweedie
family, where $\operatorname{SD}\propto\mu^{p/2}$), though the effective exponent is known to be model-dependent
\autocite{Jorgensen1994,MackVenter2000}.} Relative to the finite-variance benchmark, $k>\tfrac12$ is a shallower decline
than independent pooling gives, as co-movement between blocks would produce \autocite{Assa2025}; it does not by
itself measure that co-movement. Heavier tails can break both benchmarks, so they hold for finite
variance only (\supp{13} gives the bounding argument and an $\alpha$-stable counterexample).

\pnum Dispersion may not vanish even for the largest books, so the size law adds an undiversifiable floor in
quadrature to a diversifiable power-law term. The \emph{base size scale} is
\begin{equation}\label{eq:scale}
s(R)=\sqrt{\,\sigma_{\mathrm{undiv}}^{2}+\sigma_{\mathrm{div}}^{2}\bigl(R/R_{\mathrm{ref}}\bigr)^{2(k-1)}\,},
\qquad R_{\mathrm{ref}}=500\text{m},
\end{equation}
where $\sigma_{\mathrm{div}}$ is the diversifiable scale at the reference size and $\sigma_{\mathrm{undiv}}$ the floor.
So $k$ is the exponent of the diversifiable term, not the elasticity of the whole scale. That elasticity is
\[
  \frac{d\log s}{d\log R}=(k-1)\,w(R),
  \qquad
  w(R)=\frac{\sigma_{\mathrm{div}}^{2}(R/R_{\mathrm{ref}})^{2(k-1)}}{s(R)^{2}},
\]
where $w(R)$ is the diversifiable share of the squared base scale. The scale elasticity runs from about $k-1$ for a
small book, where $w$ is near one, toward zero for a large one, where the floor takes over. At the posterior means it is $-0.33$ at
100m, $-0.26$ at 500m and $-0.17$ at 2{,}000m, against $-\tfrac12$ throughout for square-root scaling (\supp{13}).
So when a donor larger than the target is rescaled onto a smaller book, the floor limits the amplification relative
to square-root scaling.

\pnum For the headline fit we confine $k$ to the adopted finite-variance benchmark bracket, through
$k=\tfrac12+\tfrac12\operatorname{logistic}(\theta)$ with $\theta$ on the real line. The prior therefore puts $k$ in
$(\tfrac12,1)$ before any data are seen. So $P(k>\tfrac12)$ and $P(k<1)$ are one by construction, and an interval for
$k$ inside the bracket is not evidence for or against either benchmark; under heavier tails an exponent above
$\tfrac12$ is also consistent with independent blocks. The evidence on the two ends comes from fixed-$k$ comparisons
and from refits with the lower end released (Section~\ref{sec:results:oos}).

\subsection{The fitted model}
\label{sec:framework:model}
\pnum The transfer's scale is $s(R)$ alone. But the fit has to allow for two features of the data that the size law
does not describe: some reporting years are more volatile market-wide than others, and records in the RITC regime may
have different tails. The transfer does not divide either out of a donor: the donor year's realised disturbance stays in
its movement, and an RITC donor's tail is mapped, not removed (Section~\ref{sec:framework:operator}). The model is a scale-transfer construction estimated with a Student-$t$ working likelihood. The
fitted observation scale is
\begin{equation}\label{eq:model}
\sigma_{i,t}=\exp\bigl(\eta_t+\beta_{\mathrm{RITC}}\mathbf 1_{i,t}\bigr)\,s(R_{i,t}),
\qquad \eta_t\sim\mathcal N(0,\tau),
\end{equation}
and
\begin{equation}\label{eq:like}
S_{i,t}\sim\text{Student-}t\bigl(\nu_{i,t},\,0,\,\sigma_{i,t}\bigr),
\qquad
\nu_{i,t}=\begin{cases}
\nu_{\mathrm{clean}}, & \text{clean},\\[2pt]
\nu_{\mathrm{clean}}\,e^{-\lambda_{\mathrm{RITC}}}, & \text{RITC regime},
\end{cases}
\end{equation}
where $\mathbf 1_{i,t}=1$ in the RITC regime and $s(R)$ is the base size scale of Equation~\eqref{eq:scale}. The
reporting-year effect $\eta_t$ multiplies the scale of shocks whose location is zero, so it makes some years more
volatile market-wide but cannot move the market in a common direction. We use ``scale'' for the Student-$t$ scale
throughout, never for a standard deviation.

\pnum The Student-$t$ likelihood has two jobs. It estimates the size law robustly, because the movements are
heavy-tailed and a Student-$t$ likelihood resists the outliers that would dominate a least-squares fit. And it
estimates the regime parameters that the rank map of Section~\ref{sec:framework:operator} needs. It does not supply
the transferred tail. For a clean donor and a clean target, the Student-$t$ distribution function and its inverse
cancel, and the transferred movement is $c_iS_i$ of Equation~\eqref{eq:transfer}: the observed shock at the target's
scale. So transferred quantiles are empirical quantiles of observed shocks, and the fitted shock distribution enters
only through the scales and, for RITC donors, the rank map.

\pnum The RITC tail shift is left unconstrained, $\lambda_{\mathrm{RITC}}\sim\mathcal N(0,0.7)$, so its sign is
estimated. The scale term $\beta_{\mathrm{RITC}}$ is fitted as a falsification term for treating RITC as a tail-only
effect; the operator omits it, at a cost of about $0.2\%$ of the worked stress (\supp{15}). The location is fixed at
zero. We fit the model in a Bayesian framework with weakly informative priors and hierarchical reporting-year effects
(\ref{app:notation}). Intervals on parameters are posterior credible intervals, reported as highest-density
intervals (HDI), and probabilities are posterior probabilities, with the prior's beside any that the prior alone
nearly gives. A direction is resolved when its posterior probability is at least $0.95$, and a lean when it is above
one half but short of $0.95$. The posterior is conditional on the likelihood's conditional-independence assumption:
the fit carries no cluster-aware covariance, and no transferred-stress interval corrects for dependence within a
syndicate (\supp{16}; serial dependence in \supp{17}).

\subsection{The operator and the RITC rank map}
\label{sec:framework:operator}
\pnum The full operator adds one step to Equation~\eqref{eq:transfer}. To transfer a donor movement $S_{i,t}$ onto a
target with reserves $R_q$: standardise it by the donor's base scale $s(R_{i,t})$; map it from the donor's tail
regime to the target's by a Student-$t$ quantile transform; and rescale by the target's base scale:
\begin{equation}\label{eq:operator}
S^{(q)}_{i,t} = s(R_q)\,
  F^{-1}_{\nu_q}\!\Bigl(F_{\nu_{i,t}}\bigl(S_{i,t}/s(R_{i,t})\bigr)\Bigr),
\end{equation}
where $F_\nu$ is the Student-$t$ cumulative distribution function (CDF), $\nu_{i,t}$ the donor's fitted regime and $\nu_q$ the
selected target regime. When $\nu_{i,t}=\nu_q$ the quantile transform is the identity, and
Equation~\eqref{eq:operator} reduces to the scale ratio of Equation~\eqref{eq:transfer}.

\pnum The clean target is the adopted default, and every figure below uses it. A clean donor then transfers by the
scale ratio alone. Under the clean target an RITC donor's standardised movement is rank-mapped onto the clean tail
before rescaling (the ``de-RITC'' transfer): it is read at the clean-regime quantile of the same rank. That rank is
$F_{\nu_{i,t}}$ of the base-standardised movement $S_{i,t}/s(R_{i,t})$. The input still carries the donor year's
multiplier $e^{\eta_t}$ and the fitted RITC scale multiplier $e^{\beta_{\mathrm{RITC}}}$, neither of which the
operator divides out, so the rank is a convention the operator imposes, not in general the donor's fitted conditional
percentile. The name labels the map; it does not claim that the transaction has been removed (\supp{15}). The released transfer tool also offers
an RITC-affected target and a preserve-donor-regime diagnostic, which keeps each donor's own regime so that only the
scale ratio acts.

\pnum The operator acts on the raw donor movement and uses the base scale, not the full fitted scale of
Equation~\eqref{eq:model}: it omits the RITC scale term $\beta_{\mathrm{RITC}}$ and the reporting-year effect. So no
reporting-year effect is divided out of the donor or drawn for the target. The donor year's realised disturbance is
carried across, as is any persistent level in its observed movement, scaled for a clean donor and through the rank map for an RITC
donor (\supp{16}). Every
headline figure comes from the headline fit, the model of Section~\ref{sec:framework:model} with the concentration
exponent of Section~\ref{sec:framework:concentration} fixed at zero, and uses that fit's own posterior for $k$, the
floor, the diversifiable scale and the tail indices. This headline operator, with the clean target and the
concentration exponent at zero, is the \emph{size-only transfer}.

\pnum The transferred pool is read empirically. Its quantile at level $\alpha$ is
$\text{VaR}_\alpha=\min\{x:\hat F(x)\ge\alpha\}$, one of the pool's own values, and the tail Value-at-Risk (TVaR)
is the mean at or beyond it (a tail conditional mean, not the coherent expected shortfall). The transferred empirical
99.5\% quantile is the paper's illustrative 1-in-200 stress. It is not a calibrated 1-in-200, for reasons
Section~\ref{sec:results} gives.

\subsection{The concentration overlay}
\label{sec:framework:concentration}
\pnum Two portfolios of equal size but different business mix need not have equal reserve volatility. The overlay lets
concentration modify the pooling credit through an \emph{effective size}
\begin{equation}\label{eq:reff}
R^{\mathrm{eff}}_{i,t} = R_{i,t}\,(1/H_{i,t})^{\gamma},
\qquad \gamma\ge 0,
\end{equation}
with $H$ the premium HHI of Equation~\eqref{eq:hhi}. The overlay's scale $s(R,H;\gamma)$ is Equation~\eqref{eq:scale}
with $R^{\mathrm{eff}}$ in place of $R$, and $\gamma=0$ recovers the size-only base scale. A more concentrated book
(fewer effective lines) has a smaller effective size and so less pooling credit. The prior only rules out the perverse
sign, $\gamma\ge0$. The form is motivated by a finite-variance illustration (\supp{13}); it is a phenomenological device
fitted at the portfolio level, and $H$ is a proxy for diversification opportunity, not a dependence model
\autocite{Burgi2008}.

\pnum The overlay is an exploratory sensitivity, and the data barely inform it. In the $\gamma$-free fit, the same
model with $\gamma$ estimated, $\gamma=0.394$ (95\% HDI 0.000 to 0.932) and $P(\gamma>0.05)=0.93$, against $0.96$
under the prior. It does not demonstrate an out-of-sample improvement beyond size (Section~\ref{sec:results:oos}), and its proxy is coarse
(Section~\ref{sec:data:lob}, \supp{19}). So the headline fixes $\gamma=0$, and wherever the overlay is reported it uses
the $\gamma$-free fit's own posterior and is labelled as a sensitivity. For example, take a donor of almost exactly the
500m target's size but with a more concentrated mix, Syndicate~1880 in 2024 (480.7m, $H=0.334$), and a target with
$H=0.17$. The size-only transfer leaves its movement of 0.051 unchanged in effect ($c_i=0.990$), and only the overlay
trims it, to 0.048 (\supp{22}).

\section{Evidence and limits of the transfer}
\label{sec:results}
This section asks whether size helps, and where the transfer fails. It reports the fit, the held-out density
comparison, a held-out backtest of the transferred tail, the smallest books' undercoverage and the sensitivity to
cohort depth.

\subsection{The principal fit}
\label{sec:results:posterior}
\begin{figure}[!t]
\centering
\includegraphics[width=0.75\linewidth]{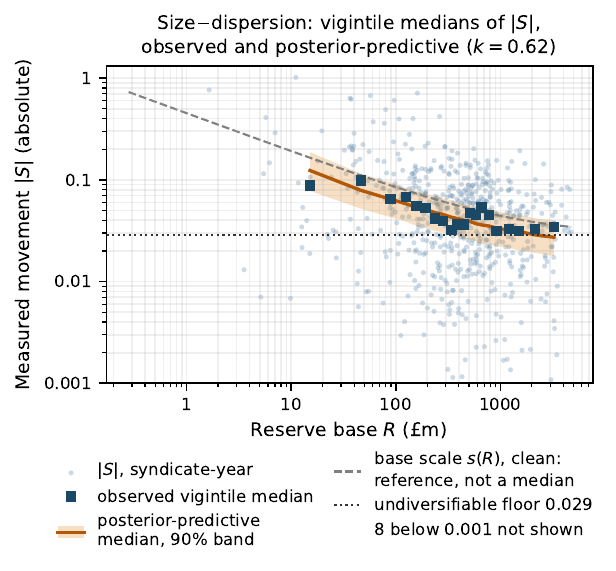}
\caption{The measured movement $|S|$ against the reserve base $R$ (log--log). Points are vigintile medians of the observed
$|S|$. The solid curve and its band are the headline fit's posterior-predictive median of $|S|$ for the same
observations, with a 90\% band. Each posterior draw simulates a new reporting year's shock with its standard
deviation $\tau$ fixed at the posterior mean, 0.137, because the fit's $\tau$ draws are not saved; the band is
conditional on that $\tau$ and does not integrate over its uncertainty. The dashed grey line is
the base scale $s(R)$ of Equation~\eqref{eq:scale} ($k=0.62$, floor $\sigma_{\mathrm{undiv}}=0.029$), a clean
base-scale reference and not a median. The vertical axis starts at $0.001$; the eight syndicate-years below it (one
with $S=0$) are not shown.}
\label{fig:size}
\end{figure}

\pnum Table~\ref{tab:posterior} gives the headline fit. The diversifiable exponent is $k\approx0.62$ (95\% HDI 0.52
to 0.73), inside the bracket the prior imposes, so the diversifiable part of the scale of $S$ decays as $R^{k-1}$,
about $R^{-0.38}$. With the floor, the base scale of $S$ on a 100m book is about $2.19$ times that on a 2{,}000m book, and that is the ratio
the size-only transfer applies between them (Figure~\ref{fig:size}).

\begin{table}[H]
\begin{threeparttable}
\caption{Posterior summary of the headline fit: the model of Equations~\eqref{eq:model} and~\eqref{eq:like} with the
concentration exponent fixed at 0 ($n=632$ syndicate-years from $J=117$ syndicates, 10 reporting years, 2015--2024).
Posterior means and 95\% highest-density intervals.}
\label{tab:posterior}
\begin{tabular}{llcc}
\toprule
\headrow Parameter & Meaning & Mean & 95\% HDI \\
\midrule
$k$                       & exponent of the diversifiable term (bracketed)  & 0.624 & [0.524, 0.735] \\
$\sigma_{\mathrm{undiv}}$ & undiversifiable floor (scale)                   & 0.029 & [0.013, 0.043] \\
$\sigma_{\mathrm{div}}$   & diversifiable scale at 500m                     & 0.044 & [0.032, 0.056] \\
$\nu_{\mathrm{clean}}$    & clean-regime degrees of freedom                 & 3.96  & [2.79, 5.30] \\
$\nu_{\mathrm{RITC}}$     & RITC-regime degrees of freedom                  & 5.39  & [1.13, 11.80] \\
\bottomrule
\end{tabular}
\begin{tablenotes}[hang]
\item[] The support of $k$ is imposed by the bracket (Section~\ref{sec:framework:size}), and the floor's existence and
the large-book asymptote are not identified by the data (see the text). Both $\sigma$ are
Student-$t$ scale parameters. The other parameters ($\lambda_{\mathrm{RITC}}$, $\beta_{\mathrm{RITC}}$, $\tau$), the
priors, sampler diagnostics and the overlay's $\gamma$-free posterior are in \ref{app:notation}.
\end{tablenotes}
\end{threeparttable}
\end{table}

\pnum The tails are heavy. The clean-regime degrees of freedom are near 4, and none of the 6{,}000 posterior draws
falls below 2, so the clean regime's variance is finite in every draw. Whether the RITC tail is heavier than the clean
one is not settled: the posterior probability that it is heavier is 0.40, against 0.50 under the prior. Across four RITC treatments the size
exponent ($0.613$--$0.624$), the floor ($0.029$--$0.030$) and the clean tail index barely move (\supp{15}). The clean
base-scale residuals lean adverse in the far tail (\ref{app:robust:fit}); a skew-$t$ refit moves the stress by less
than the decision line set before it (\supp{20}), so the headline keeps the symmetric shock.

\pnum The floor is a structural choice, not a finding. Above about 1bn the conditional size trend is not resolved, so further
size credit there can be neither demonstrated nor ruled out (\supp{14}). A floorless law extrapolates toward zero
volatility as a fraction of reserves for books larger than any in this market. A floor, if the decline does continue,
overstates a very large book's volatility by a margin that grows with its size. We retain a positive asymptote as a
prudential extrapolation that these data do not identify. Its level trades off against the exponent
(\ref{app:robust:floor}). A syndicate random intercept lowers it to 0.016 (\supp{16}),
while the fast run-off maturity rebasing raises it to 0.081 (\supp{4}); the latter uses a proxy for the reserve split.

\subsection{Density prediction for unseen syndicates}
\label{sec:results:oos}
\pnum The paper uses three validation designs, and they answer different questions:
\begin{enumerate}
  \item \emph{five-fold syndicate cross-validation} scores a fitted density: whole syndicates are held out, the model
        is refitted on the rest, and the held-out movements are scored by expected log pointwise predictive density
        (ELPD), and each score difference is read through a Bayesian bootstrap over syndicates; it tests the scale
        specification (Table~\ref{tab:oos});
  \item a \emph{ten-fold syndicate backtest} scores the transferred empirical quantiles against held-out movements,
        with the donors drawn from the training folds only (Section~\ref{sec:results:backtest});
  \item leaving the target's own reporting year out of the donor pool changes the donor library, not the fit, so it
        is a sensitivity of the backtest and not a chronological test (\supp{21}).
\end{enumerate}
None of them forecasts a firm's next year from the information available at the time.

\pnum Size predicts unseen syndicates far better than a single market scale (Table~\ref{tab:oos}, Panel A). The
single-regime size-only model's held-out ELPD exceeds the naive pool's by $+34.4$ (95\% credible interval $+11$ to
$+64$; posterior probability 0.998 that size-only predicts better). That is $0.054$ per held-out syndicate-year: a
geometric-mean predictive density about $6\%$ higher. The constant-scale specification ($k=1$) is
strongly disfavoured by held-out density prediction: fitted separately, it scores within 0.1 of the naive pool. A
marginal size law does not identify the joint dependence between the blocks of a book, so this tells against a
constant scale, not against every form of comonotonic dependence. The adopted two-regime model, scored as the paper uses it (a held-out
syndicate in a new reporting year, with the year effect integrated out), ties the single-regime size-only model. The
central 90\%, 95\% and 99\% plug-in prediction intervals of the single-regime composition model are close to nominal out of
sample (Table~\ref{tab:ooscal}), but that check does not test the one-sided 99.5\% tail. Nor does a better
held-out density show that the transferred 99.5\% quantile is calibrated; that is the backtest's question
(Section~\ref{sec:results:backtest}).

\begin{table}[H]
\begin{threeparttable}
\caption{Five-fold cross-validation by syndicate (no syndicate in both training and test), scored by held-out ELPD.
Panel A compares the headline's size law with a single market scale; Panel B scores secondary specifications.}
\label{tab:oos}
\begin{tabular}{lcc}
\toprule
\headrow Model & Held-out ELPD & $\Delta$ELPD vs.\ naive \\
\midrule
\multicolumn{3}{l}{\textit{Panel A. Size against a single market scale}} \\
Naive market pool (one scale for every book) & 667.7 & -- \\
Size-only, single regime (size $+$ floor) & 702.0 & +34.4 \\
Adopted two-regime size-only (year effect integrated out) & 701.0 & +33.4 \\
\midrule
\multicolumn{3}{l}{\textit{Panel B. Secondary specifications}} \\
Composition (size $+$ concentration $+$ floor) & 701.1 & +33.4 \\
Fixed $k=\tfrac12$ (with concentration and floor) & 699.8 & +32.2 \\
Constant scale, $k=1$ (fitted separately) & 667.8 & +0.1 \\
\bottomrule
\end{tabular}
\begin{tablenotes}[hang]
\item[] Each $\Delta$ELPD is computed from unrounded scores, so it can differ by 0.1 from the difference of the
rounded columns. All rows except the adopted model are fitted single-regime and without
the reporting-year effect. Intervals
are Bayesian-bootstrap 95\% credible intervals over syndicates: size-only minus naive $[+11,+64]$; adopted minus naive
$[+9,+64]$, and adopted minus single-regime size-only $-1.0$ $[-3.7,+2.4]$; composition minus naive $[+10,+63]$. The
fixed-$k=\tfrac12$ row carries the concentration term, as the composition row does; no size-only density fit with
$k=\tfrac12$ exists. Square-root scaling without a floor is a different object, scored in Table~\ref{tab:backtest}.
\end{tablenotes}
\end{threeparttable}
\end{table}

\pnum Panel B adds little. No improvement over size is established: the composition model is 1.0 ELPD behind size-only
(posterior probability 0.24 that it predicts better). Nor do the data separate the free exponent from $k=\tfrac12$:
with the concentration term carried, the free exponent leads by $+1.24$ ($-4.0$ to $+6.2$; posterior probability
0.70), and fixing $k=\tfrac12$ lowers the worked stress of Section~\ref{sec:vignettes} from 0.342 to 0.315. The bracket
is not what holds $k$ above $\tfrac12$: releasing its lower end gives $P(k<\tfrac12)=0.07$, against 0.32 and 0.50
under the two priors (\ref{app:robust:pooling}). So the data establish that dispersion falls with size. The exact exponent is unresolved: slower-than-independent pooling is neither shown nor
excluded.

\subsection{A held-out backtest of the transferred tail}
\label{sec:results:backtest}
\pnum Table~\ref{tab:backtest} scores the transferred quantiles themselves. Syndicates were assigned to ten folds.
For each fold the headline model was refitted on the other nine, and each held-out observation was compared with the
transferred empirical quantile of the donors from those nine training folds, transferred to its own reserves at the
fold-fit posterior-mean parameters. A whole fold is left out of both the fit and the pool. For example, 1991/2020 is scored against 566 donors, because its whole
66-record fold is left out, not only Syndicate~1991's other years. The raw pool and square-root scaling of the same
pool face the same test. Each set of transferred quantiles is also scored by its mean pinball (quantile)
loss: a forecast $q$ of the level-$\alpha$ quantile scores $\alpha(y-q)$ for an outcome $y\ge q$ and
$(1-\alpha)(q-y)$ for $y<q$, averaged over the targets. Lower is better, and it penalises both a miss and a
quantile set needlessly high. A separate leave-one-syndicate-out calculation that reuses the full-sample parameters is a
different design, reported in \supp{21} and not used here.

\pnum Exceedances are counted overall and by quartile of the target's reserves. The quartiles are the primary
sample's own, with boundaries 172m, 421m and 816m, and they describe the grouping only. Each count is read through a
Bayesian bootstrap over syndicates, which gives a posterior for the exceedance rate $r$ and the posterior probability
that $r$ exceeds the nominal rate.

\begin{table}[H]
\begin{threeparttable}
\caption{Ten-fold syndicate backtest of the transferred empirical quantiles: exceedances of the 99\% and 99.5\%
quantiles by held-out syndicate-years, overall and by quartile (Q1 to Q4) of the target's reserve base, for the
raw pool, square-root scaling and the headline size-only transfer.}
\label{tab:backtest}
\small
\setlength{\tabcolsep}{3.5pt}
\begin{tabular}{llccccc>{\centering\arraybackslash}p{1.9cm}}
\toprule
\headrow $\alpha$ & Method & All & Q1 & Q2 & Q3 & Q4 & All-target mean pinball loss ($\times10^{-3}$) \\
\midrule
Targets / synd. & & 632 / 117 & 158 / 63 & 158 / 53 & 158 / 42 & 158 / 28 & -- \\
\midrule
99\% & expected & 6.3 & 1.6 & 1.6 & 1.6 & 1.6 & -- \\
 & raw pool & 9 & 9 & 0 & 0 & 0 & 9.2 \\
 & $\sqrt R$ & 7 & 2 & 1 & 0 & 4 & 4.5 \\
 & operator & 10 & 9 & 1 & 0 & 0 & 5.8 \\
 & synd.\ with a miss (op.) & 5 & 4 & 1 & 0 & 0 & -- \\
\midrule
99.5\% & expected & 3.2 & 0.8 & 0.8 & 0.8 & 0.8 & -- \\
 & raw pool & 5 & 5 & 0 & 0 & 0 & 4.5 \\
 & $\sqrt R$ & 3 & 1 & 0 & 0 & 2 & 2.7 \\
 & operator & 7 & 6 & 1 & 0 & 0 & 2.9 \\
 & synd.\ with a miss (op.) & 4 & 3 & 1 & 0 & 0 & -- \\
\bottomrule
\end{tabular}
\begin{tablenotes}[hang]
\item[] synd.: distinct syndicates; operator (op.): the headline size-only transfer. The row synd.\ with a miss counts the
distinct syndicates among the transfer's exceedances; it is not a count of exceedances. Counts use forecasts from
models refitted within each held-out fold: each of the 632 targets is scored against donors from the nine training
folds. Square-root scaling ($\sqrt R$) multiplies each donor by $\sqrt{R_i/R_q}$, with no floor and no rank map. An
RITC-regime target is scored under the clean target regime. Expected counts are $1-\alpha$ times the targets in each
column. The last column is the mean pinball loss over all 632 targets; the 90\% and 95\% levels, the fixed original
size bands and the full-sample-parameter version are in \supp{21}.
\end{tablenotes}
\end{threeparttable}
\end{table}

\pnum Square-root scaling has a descriptively lower mean pinball loss in the extreme tail. Over all targets it is lower than the
size-only transfer's at 99\% (4.5 against 5.8, $\times10^{-3}$) and at 99.5\% (2.7 against 2.9). At 95\% the size-only
transfer is lower by a negligible margin (13.9 against 14.0), and at 90\% by a larger observed margin (20.1 against 21.8). We have not put an
interval on these paired differences, so the comparison is descriptive. Square-root scaling has fewer small-book misses
and more among the largest books: at 99.5\% it has one in the smallest quartile and two in the largest, against the
size-only transfer's six and none. At 99\% its advantage sits in the smallest quartile, where its mean pinball loss is
10.2 against 15.3; in the three larger quartiles the two tie by pinball loss, though square-root scaling has five
exceedances there against one. The raw pool is the worst at every level. A size-banded benchmark, which reads the tail
of the donors whose reserves lie within a factor of two of the target's without rescaling, has too few donors at the
99.5\% point to say which tail is right (\supp{21}).

\pnum So the fitted exponent and floor have no demonstrated extreme-tail advantage over the simple benchmark. What the
fitted law adds is an estimated size dependence with explicit assumptions. Transfer by named donor and inspectable donor
support belong to the framework: a named-donor square-root transfer would have them too. We did not select square-root scaling on these outcomes and do not offer it as
a replacement. A fixed exponent or deleted floor in the headline transfer, or a minimum tail multiplier, would need
its own refit and held-out tail assessment; we have not made those comparisons.

\subsection{Undercoverage in the smallest books}
\label{sec:results:smallbook}
\pnum In the smallest reserve-size quartile, six of 158 observations exceeded the transferred 99.5th percentile,
against 0.79 expected under nominal coverage. That is a rate of 3.8\%, against 0.5\%. At the 99\% level the count is
nine, against 1.58 expected.

\pnum The misses are clustered. The six 99.5\% events come from three syndicates: Syndicate~1991 in 2015, 2016, 2019
and 2020, Syndicate~1945 in 2020, and Syndicate~5820 in 2016. At 99\% they are joined by 1991/2017, 1991/2018 and
3334/2016, so Syndicate~1991 supplies six of the nine. The by-syndicate bootstrap allows for this clustering. It gives
the 99.5\% rate a posterior mean of 3.7\% (95\% credible interval 0.7\% to 9.8\%), with posterior probability 0.988 that
it exceeds 0.5\%; at 99\% the mean is 5.6\% (1.2\% to 14.9\%), with probability 0.991 that it exceeds 1\%. These
probabilities are for this quartile alone; they are not adjusted for reading four quartiles at two levels.

\pnum The higher quartiles do not show the same excess of misses, but the sparse extreme events do not establish
99.5\% calibration there. Across the three of them there is one miss at each level, 3010/2024, a second-quartile book, against 4.7
expected at 99\% and 2.4 at 99.5\%. So no excess of misses was observed there. With counts this sparse, that does not rule out under-coverage at
99.5\%, and a one-sided count cannot tell over-coverage from chance.

\pnum The source evidence identifies mechanisms that may contribute to these misses (Table~\ref{tab:tailcases}). In
each small-book event syndicate the filings describe exposure still earning, or old underwriting years still carrying
exposure, so an age cutoff did not isolate re-estimation of opening reserves. Each also reports genuine adverse
experience. The accounts do not separate the amounts, so we subtract no estimated earning component, and these
mechanisms do not explain the misses away. For 3010/2024, the gross increase is concentrated in one older cohort,
while the same cohort's net movement is favourable; the filing does not establish the earning share or cause (\supp{1}).

\begin{table}[!htbp]
\begin{threeparttable}
\caption{The held-out misses and what the filings disclose about them. Exact quotations, pages and arithmetic are in
\supp{1}.}
\label{tab:tailcases}
\footnotesize
\begin{tabular}{>{\raggedright\arraybackslash}p{0.12\linewidth}>{\raggedright\arraybackslash}p{0.16\linewidth}>{\raggedright\arraybackslash}p{0.35\linewidth}>{\raggedright\arraybackslash}p{0.23\linewidth}}
\toprule
\headrow Record & Misses & Disclosed feature & What remains unknown \\
\midrule
1991, 2015--2020 (Q1) & 99.5\%: 2015, 2016, 2019, 2020; 99\%: also 2017, 2018 & 2020 triangle on an earned basis;
delayed inception and late coverholder attachments, with exposure continuing into later years; genuine adverse and coronavirus disease (COVID)
experience in 2020 & How much of each movement is earning emergence \\
5820, 2016 (Q1) & 99.5\% & Business earning over five years; triangles earned to date; unearned premium in older
cohorts; separately reported net deterioration & The split of the gross movement between earning and deterioration \\
1945, 2020 (Q1) & 99.5\% & COVID losses on discontinued Contingency business attached to older years of account, with
exposure to 2021 events & The earning share; the calendar release and the cohort increase are different populations \\
3334, 2016 (Q1) & 99\%; 99.5\% when 1991 is omitted & Prolonged earning on the 2014 cohort; adverse older liability
experience & The split between the two \\
3010, 2024 (Q2) & 99\% and 99.5\% & Gross increase concentrated in one cohort; net movement in that cohort favourable & Its cause and earning share \\
\bottomrule
\end{tabular}
\end{threeparttable}
\end{table}

\FloatBarrier

\pnum Two refits measure how much of the result is influence (Table~\ref{tab:samplesens}). One omits the six 1991
years; the other adds back the 11 set-aside records at their pre-census figures. On the fixed band, omitting 1991
halves the misses, from 6 of 159 to 3 of 153. On the common targets, the refit and rebuilt pool raise the
small-band 99.5\% misses from 2 to 3, the 99\% count stays at 3, and 3334/2016 becomes a
99.5\% miss. Deleting a syndicate removes both its adverse targets and its severe donors, so a lower miss count is not
improved prediction. Source eligibility (Section~\ref{sec:data:filters}) decides the primary sample; these refits
measure influence only (\supp{21}).

\begin{table}[H]
\begin{threeparttable}
\caption{Influence of the sample on the small-book misses: three samples on the original fixed size bands (Panel A),
and the same three fits scored on common targets (Panel B).}
\label{tab:samplesens}
\small
\begin{tabular}{lccc}
\toprule
\multicolumn{4}{l}{\textit{Panel A. Each sample's own targets, smallest fixed band (below 173m), 99.5\%}} \\
\headrow Sample & Records / syndicates & Band misses & Worked stress \\
\midrule
Primary & 632 / 117 & 6 of 159 & 0.342 \\
Comparator: set-aside records added back & 643 / 118 & 6 of 161 & 0.336 \\
All six 1991 years omitted & 626 / 116 & 3 of 153 & 0.293 \\
\midrule
\multicolumn{4}{l}{\textit{Panel B. 626 common targets (153 in the band), smallest-band misses}} \\
\headrow Fits and donor pools & 99\% & 99.5\% & \\
\midrule
Primary & 3 & 2 & \\
Comparator & 3 & 2 & \\
1991 omitted & 3 & 3 & \\
\bottomrule
\end{tabular}
\begin{tablenotes}[hang]
\item[] Every sample is refitted on its own ten folds and its donor pools rebuilt. The fixed band is the earlier
sample's smallest quartile, held fixed so that the groupings are comparable; it differs from the primary's own
quartile of Table~\ref{tab:backtest}. The common targets exclude the 11 set-aside records and all six 1991 years, so
Panel B measures 1991's effect on the forecasts of other targets only. Worked stress: the transferred 99.5\% quantile
for the target of Section~\ref{sec:vignettes}.
\end{tablenotes}
\end{threeparttable}
\end{table}

\subsection{Cohort depth}
\label{sec:results:depth}
\pnum The primary sums the movement over underwriting years up to $t-2$. The youngest of them, $u=t-2$, is in its
third year of development, when a Lloyd's year of account normally closes, so it can still carry early development and
earning. A sensitivity moves the cut to $u\le t-3$, recomputing each figure from a stored grid or, where needed, source-read register cells. Of the 632 primary
records, 44 have no cohort up to $t-3$, and 33 of those are in the smallest quartile, mostly young small books. So the
comparison is made on the other 588 records, from 104 syndicates: a selected sample.

\pnum The measured movement is materially sensitive to its youngest included cohort. On the 588 records, moving the
cut from $u\le t-2$ to $u\le t-3$ lowers the worked stress of Section~\ref{sec:vignettes} by 24.2\%, to 0.262 (both
depths fitted on the same records), and the fitted law moves with it: $k$ rises from 0.60 to 0.74 and the floor falls
from 0.030 to 0.018. The same 0.262 is 23.4\% below the full primary's 0.342; that is a different comparison, across
populations. In the refitted backtest on the 588 records, whose own quartiles hold 147 targets each, the smallest
quartile's 99\% misses fall from 8 to 3, against 1.5 expected, and the posterior probability that the rate exceeds 1\%
falls from 0.99 to 0.82. Over all 588 targets they fall only from 9 to 7, against 5.9 expected, because the two largest
quartiles gain three misses (\supp{3}). But fewer misses on a selected sample are not a demonstrated cure for the small-book tail: the 44 records
the deeper cut cannot measure include one of the primary's small-book misses, 1991/2015. We keep $u\le t-2$, as
Equation~\eqref{eq:severity} defines it, because the deeper cut cannot be measured for a selected group of small books.

\section{A worked target and its uncertainty}
\label{sec:vignettes}
This section transfers the whole primary sample onto one illustrative target and reads its stress at the precision shown.

\subsection{From the raw pool to the target basis}
\label{sec:vignettes:entrant}
\pnum The target is a \pounds500m portfolio with no usable history of its own: for example, a syndicate that acquires a
mature book of that size. (A syndicate that starts from nothing has no older cohorts to stress.) We transfer all 632
donor movements onto its basis with the size-only transfer. Donor~A of Section~\ref{sec:framework:transfer} is one of
them, and becomes 0.083.

\pnum Table~\ref{tab:worked_example} compares three ways of reading the target's adverse tail: the raw pool takes the
donors' movements as they are, square-root scaling multiplies each by $\sqrt{R_i/500}$, and the size-only transfer
applies the fitted law. The transferred empirical 99.5\% quantile falls from 0.672 in the raw pool to 0.342 under the
fitted transfer, and square-root scaling gives 0.304. In both the raw and the transferred pool the 99.5\% quantile is
the same donor-year, 5820/2016, the opening example of Section~\ref{sec:intro}. The reduction holds across the
95--99.5\% tail, in the tail conditional mean, and under a generalised-Pareto cross-check (\supp{23}); the pool's
standard deviation falls from 0.112 to 0.074.

\begin{table}[H]
\begin{threeparttable}
\caption{The \pounds500m target's adverse tail. Panel A: three readings of the transferred pool. Panel B: the donors
at or beyond the size-only transfer's 99.5\% quantile.}
\label{tab:worked_example}
\label{tab:v1donors}
\small
\textit{Panel A. The pool's adverse tail}\\[2pt]
\begin{tabular}{lccc}
\toprule
\headrow Measure & Raw pool & Square-root scaling & Size-only transfer \\
\midrule
$\text{VaR}_{99}$ & 0.605 & 0.250 & 0.276 \\
$\text{VaR}_{99.5}$ & 0.672 & 0.304 & 0.342 \\
$\text{TVaR}_{99}$ & 0.718 & 0.308 & 0.345 \\
\bottomrule
\end{tabular}

\vspace{4pt}
\textit{Panel B. Donors at or beyond the 99.5\% quantile}\\[2pt]
\begin{tabular}{rlrrrrl}
\toprule
\headrow Rank & Donor & Raw $S$ & $R$\,(m) & $c_i$ & $S^{(q)}$ & Source basis \\
\midrule
1 & 1991/2020 & 0.711 & 161 & 0.719 & 0.511 & earned basis \\
2 & 1945/2020 & 0.605 & 97 & 0.609 & 0.368 & continuing old-year exposure \\
3 & 3010/2024 & 0.419 & 263 & 0.835 & 0.350 & earning unresolved \\
4 & 5820/2016 & 0.672 & 57 & 0.509 & 0.342 & long earning \\
\bottomrule
\end{tabular}
\begin{tablenotes}[hang]
\item[] Point estimates at posterior-mean parameters, with the RITC rank map applied. The 95\% interval of the size-only
$\text{VaR}_{99.5}$ is 0.234 to 0.516 (Section~\ref{sec:vignettes:uncertainty}). All four donors are clean; the
source basis is the filing feature recorded in \supp{1}; ``earning unresolved'' means the filing does not establish
the earning share or cause. The ten largest transferred donors, their posterior intervals, their ranks in a clean-only pool and the
overlay and $k=\tfrac12$ alternatives are in \supp{22}.
\end{tablenotes}
\end{threeparttable}
\end{table}

\pnum So the fitted transfer gives a lower empirical stress for this target than the raw pool: about 0.34 of reserves
against 0.67. The quantile is set by the whole ranked pool, not by how many donors are smaller than the target: a few
large, severe donors scaled up could set a higher quantile. Here they do not. The cutpoint donor in both pools is
5820/2016, on reserves of 57m; the size law multiplies its movement by about 0.509 on a \pounds500m book, so its
0.672 becomes 0.342, and every donor at or beyond it is a smaller book scaled down (Panel B). This is a
consequence of the fitted model, not an independent test of it, and it does not show that 0.34 is the 1-in-200 of
every \pounds500m book.

\begin{figure}[H]
\centering
\includegraphics[width=0.78\linewidth]{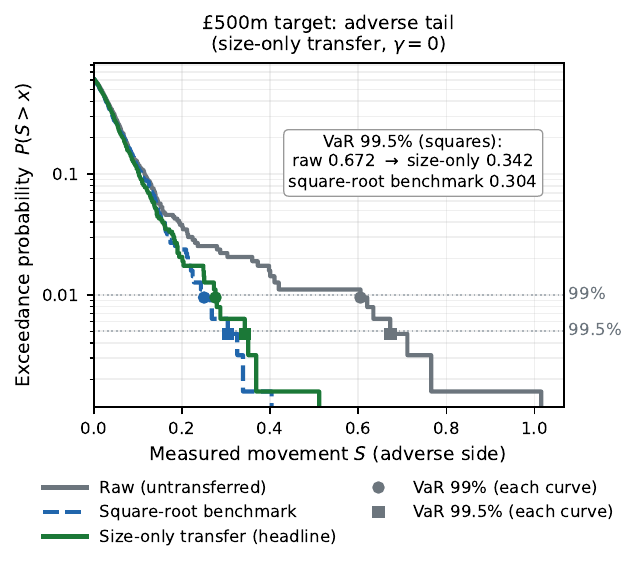}
\caption{The \pounds500m target's adverse tail: empirical survivor function $P(S>x)$ (log vertical axis) of the
transferred pool, with the 99\% and 99.5\% quantiles marked, for the raw pool (99.5\% quantile 0.672), square-root
scaling (0.304) and the size-only transfer (0.342). The size-only curve applies the RITC rank map. Carrying the 27
RITC-regime donors without the rank map gives the same 0.342: in both pools, no RITC-regime donor lies at or beyond
the 99.5\% level. Markers sit at the empirical strict-exceedance rate of each quantile; the dotted lines mark the
nominal rates. In the legend, ``Raw (untransferred)'' is the raw pool, ``Square-root benchmark'' is square-root scaling,
and $\gamma=0$ in the title marks the size-only transfer.}
\label{fig:vignette1}
\end{figure}

\pnum The 99.5\% quantile is an extreme order statistic, so we show the donors that support it. Four donor-years lie at
or beyond it (Table~\ref{tab:worked_example}, Panel B). Three come from the small-book event syndicates of
Section~\ref{sec:results:smallbook}, and the cutpoint donor is 5820/2016, so the target's stress inherits their
earning and exposure meaning. The remaining donor, 3010/2024, is the higher-quartile miss: its gross increase
is concentrated in one cohort, whose net movement is favourable, and its earning share is unresolved (\supp{1}). None of the four is in the RITC regime, so the rank map changes nothing at this point
(Figure~\ref{fig:vignette1}).

\pnum The stress rests materially on its most adverse donor, 1991/2020. Refitting without that record lowers the
plug-in transferred 99.5\% quantile from about 0.34 to 0.29 (0.342 to 0.285, about 17\%). The record passes the source
rule, so it stays; its syndicate went into run-off on 6~November 2020, part way through the year, so the year is not a
run-off year under the sample rules (\ref{app:audit:exclusions}). Removal with the fit held fixed, removal of two
donors and the posterior-mean comparisons are in \supp{22}. The very top of the transferred tail rests on few donors,
and it is sensitive to extrapolating the size law across large size gaps.

\FloatBarrier
\subsection{Uncertainty beside the stress}
\label{sec:vignettes:uncertainty}
\begin{figure}[H]
\centering
\includegraphics[width=0.95\linewidth]{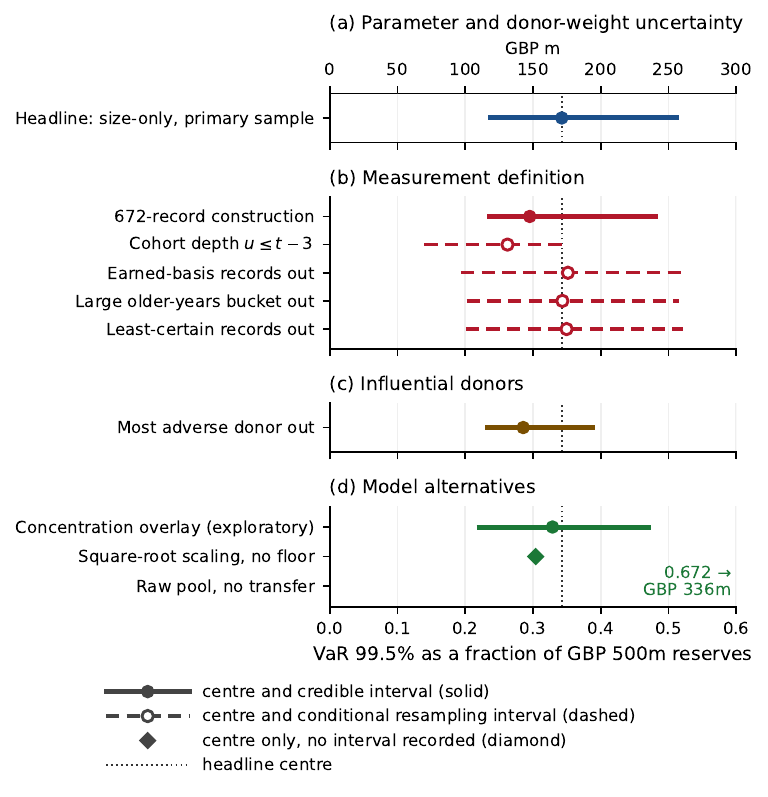}
\caption{Sources of uncertainty for the \pounds500m target's transferred 99.5\% quantile under the size-only transfer,
as a fraction of reserves: (a) parameters and donor weights; (b) measurement definition; (c) influential donors;
(d) model alternatives. A solid bar with a filled circle is a 95\% posterior interval and its centre; a long-dashed
bar with an open circle is the conditional resampling range of a refit whose record holds no posterior interval; a
diamond is a point estimate with no interval. The dotted line marks the full-sample size-only point estimate; the
cohort-depth comparison in the text uses the baseline fitted to the same 588 records. The raw pool (0.672)
lies beyond the axis. No row is summed with another.}
\label{fig:uncertainty}
\end{figure}

\pnum Figure~\ref{fig:uncertainty} keeps the sources of uncertainty apart, and they are not added into one band.
Each row answers a different question:
\begin{enumerate}
  \item[(a)] \emph{parameters and donor weights}: a 95\% posterior interval for the stress, 0.234 to 0.516 of reserves;
  \item[(b)] \emph{measurement definition}: refits on a changed outcome or sample (the 672-record construction, the
        cohort-depth cut, the earned-basis exclusion, the exclusion of records with a large older-years bucket and the
        exclusion of records the readings could not settle);
  \item[(c)] \emph{influential donors}: the stress without its most adverse donor;
  \item[(d)] \emph{model alternatives}: the concentration overlay, square-root scaling and the raw pool.
\end{enumerate}
Rows (b) to (d) are refits or alternative models, each conditional on its own sample or model; an interval drawn
on one of them is that refit's own, and none is combined with row (a). The largest moves come from the measurement
and the donors: the earlier 672-record construction gives a stress 14\% lower, the deeper cohort cut lowers the stress by 24\% on the 588 records both cuts can measure, a selected
sample that leaves out 33 records of the smallest quartile (Section~\ref{sec:results:depth}),
and removing the most adverse donor lowers it by 17\%; fixing $k=\tfrac12$ lowers it by 8\%
(Section~\ref{sec:results:oos}).
Refits weighted by an inverse response propensity, floored at three levels, toward the provisional target of
\ref{app:audit:exclusions} raise the \pounds500m stress by $2.8\%$ to $6.6\%$ at 99.5\% and by $2.6\%$ to $13.6\%$
at 99\%; they are conditional on the fitted weights and are not plotted (\supp{6}).

\pnum The interval in row (a) treats the donor pool as a sample of $J=117$ syndicates from a wider population.
That pool is the eligible sample, in which small syndicates are under-represented
(Section~\ref{sec:data:filters}), so the wider population is one of eligible, mostly established books. Each
replicate draws syndicate weights $w\sim\mathrm{Dirichlet}(1,\dots,1)$ \autocite{Rubin1981} and turns them into
scenario shares
\[
  q_s=\frac{n_s\,w_s}{\sum_{t=1}^{J} n_t\,w_t},
\]
where $n_s$ is the number of years syndicate $s$ contributes; a scenario is then a uniform draw among that syndicate's
$n_s$ years. So longer histories carry more predictive mass: at equal $w$, a syndicate with three years and one with a
single year have shares $3/4$ and $1/4$, not $1/2$ each. Each replicate pairs its weights with one draw from the fitted
posterior, every parameter read at the same index, and reads the stress from the weighted pool. The interval is the
2.5--97.5\% range over 4{,}000 replicates (\supp{22}), and the stress is below the raw pool's in 3{,}998 of them.

\pnum This quantifies which donor syndicates enter and the parameter uncertainty within one fixed model. It does not
refit a cluster-dependent likelihood or repair the posterior's conditional-independence assumption, so the interval is
conditional on the data, the model and that treatment of dependence.

\section{Applying the evidence}
\label{sec:implications}
This section says what the method lets an actuary do, what the evidence establishes, where the method fails or is
untested, and what a particular application needs.

\subsection{What the method allows, and what the evidence shows}
\label{sec:implications:practice}
\pnum The method lets an actuary put a named historical movement on a stated target basis and show the arithmetic.
For example, a syndicate challenging its internal-model 1-in-200 can place the largest adverse peer movements of the
last decade on its own size basis, list them with their sources, and ask whether its modelled stress still exceeds
them. For a clean donor each transfer is one ratio of fitted scales, so it can be shown line by line, the transparency
that professional actuarial standards require of a reserve-risk calibration \autocite{TAS100,TAS200}. The same steps
serve a scenario library, a stress set or an own risk and solvency assessment (ORSA) challenge to one-year assumptions.

\pnum The evidence establishes that the dispersion of the measured movement varies materially with size: a
size-based scale predicts held-out syndicates far better than a single market scale, the constant-scale specification
is strongly disfavoured, and the tails are heavy (Sections~\ref{sec:results:posterior} and~\ref{sec:results:oos}). It
leaves four things unresolved: the exact exponent, the level of the floor, the size of any concentration effect, and
whether the RITC tail is heavier than the clean one.

\pnum The transferred tail fails, or is untested, in four places. It understates the smallest books: six misses at
99.5\% among 158 smallest-quartile targets, from three syndicates (Section~\ref{sec:results:smallbook}). In the larger
books the sparse extreme events do not establish 99.5\% calibration either, and square-root scaling has descriptively lower
mean pinball loss at 99\% and 99.5\% (Section~\ref{sec:results:backtest}). Above about 1bn the conditional size trend is not resolved, and
beyond about 5bn every specification is an extrapolation (\ref{app:robust:floor}). And forecasting a
firm's next year is untested.

\pnum For the \pounds500m target the size law lowers the transferred stress, because the donor at its 99.5\% cutpoint
is a much smaller book. A target larger than most donors need not get a lower stress: its quantile can be set by a few
large, severe donors that the law scales up. For a book that has shrunk, the same donor pool is rescaled to a smaller
target, where the fitted base scale is larger, so every donor's multiplier rises by the same factor and the law raises
the stress as a share of reserves, though the stress in pounds falls (\supp{24}). Both are
conditional model outputs.
So a diversification credit or a reserve-risk load is an argument to be evidenced, not a number to be booked: the
monotone law that produces its direction does not validate it.

\subsection{What an application needs}
\label{sec:implications:apply}
\pnum Table~\ref{tab:use} sets out three uses, what the evidence supports for each, and what an application has to
add.

\begin{table}[H]
\begin{threeparttable}
\caption{Three uses of a transferred scenario: what the evidence supports, and what the application must add.}
\label{tab:use}
\footnotesize
\begin{tabular}{p{0.24\linewidth}p{0.33\linewidth}p{0.33\linewidth}}
\toprule
\headrow Use & What the evidence supports & What the application must add \\
\midrule
Comparing a named scenario with the target's assumptions & A size-adjusted transfer, with the scale ratio and regime map where needed, shown donor by donor
(Sections~\ref{sec:framework:transfer} and~\ref{sec:results:oos}) & The donor's measure, remaining exposure, gross
reserve base and business \\
Reading a transferred quantile as a stress & An illustration with a conditional interval and named donor support
(Section~\ref{sec:vignettes}); no demonstrated 99.5\% calibration & Evidence on calibration for the target's size; own
support for a small book (Section~\ref{sec:results:smallbook}); the influence of the top donors \\
Taking a credit or a load from size or mix & A direction that is a model output; a weakly identified concentration
overlay (Section~\ref{sec:framework:concentration}) & Evidence specific to the portfolio; above about 1bn, a justified
extrapolation (Section~\ref{sec:results:posterior}) \\
\bottomrule
\end{tabular}
\end{threeparttable}
\end{table}

\pnum Three cautions apply to every use:
\begin{enumerate}
  \item inspect each supporting donor's source basis, the exposure it still carries and the business that generated
        its reserves; this matters most for young, rapidly changing, long-earning or discontinued books, size-only
        rescaling does not adjust for it, and the absence of an earning disclosure does not validate a donor;
  \item the transferred pool carries each donor's observed movement, including any persistent average level (an RITC donor's through the rank map); such a level is
        treated here as reserving prudence, an
        interpretation the data are consistent with but do not identify; for the 7 of the 79 syndicates with at least
        three observations whose estimated means have 95\% empirical-Bayes intervals above zero, the transferred
        level and spread may be misstated,
        and a firm that wants its own level in the stress must centre the pool first, since adding a target intercept
        to an uncentred pool counts location twice (\supp{16});
  \item the transfer is fitted and applied gross of outward reinsurance, and must not be applied to net reserves
        directly.
\end{enumerate}

\subsection{Limitations and future work}
\label{sec:implications:limits}
\pnum The limitations fall under four themes, each discussed where it arises:
\begin{enumerate}
  \item \emph{measurement and source assurance}: the outcome is observed cohort development, not opening-reserve
        deterioration (Sections~\ref{sec:data:pyd} and~\ref{sec:results:smallbook}); the source rule is post hoc, the
        automated readings check fidelity to the filing, not actuarial judgement (Section~\ref{sec:data:filters}), and
        the transfer and take-on flags do not measure the transactions (\supp{9});
  \item \emph{selection and sparse tails}: small syndicates are under-represented; selection-weighted refits raise the
        \pounds500m stress by up to $6.6\%$ at 99.5\% and $13.6\%$ at 99\%, conditional on fitted weights and the provisional target (\supp{6}). The 99.5\% point
        rests on a few donors, one of which moves it by about 17\% (Section~\ref{sec:vignettes:entrant}). In a hypothetical
        target of 742, assuming 52 additional screened filings are eligible and their whole-account or transfer amounts
        proxy compatible reserve bases, a ledger-order pseudo-outcome stress moves $k$ from $0.634$ to $0.518$ as the
        pseudo-outcome scale multiplier $c$ rises from $1$ to $5$ (\supp{6});
  \item \emph{model, dependence and extrapolation}: an interval on a parameter or a transferred stress inherits the
        likelihood's conditional-independence assumption (Section~\ref{sec:framework:model}); the by-syndicate
        bootstrap of the backtest's exceedance rates shares one weight across a syndicate's years and is conditional on
        the fixed held-out forecasts, and it neither repairs the fitted parameters' covariance nor carries refit
        uncertainty (Section~\ref{sec:results:backtest}); the floor is a structural choice, with few books above 2bn
        (\ref{app:robust:floor}); and size and mix are the only drivers modelled, in one market and gross
        of reinsurance (\supp{18});
  \item \emph{evaluation scope}: the validation holds out syndicates, not years, so forecasting is untested
        (Section~\ref{sec:results:oos}).
\end{enumerate}
Future work could develop an estimand that adjusts for earning and exposure, a joint model of location and dependence,
and a chronological validation specified before its outcomes are seen.

\vspace{0pt plus 8\baselineskip}\penalty-200\vspace{0pt plus -8\baselineskip}
\section{Conclusion}
\label{sec:conclusion}
\subsection{What the evidence shows}
\label{sec:conclusion:evidence}
\pnum Historical market reserve movements can inform a portfolio's stress scenarios once their source basis and size
are made explicit. Using public Lloyd's accounts, we estimate how the dispersion of observed older-cohort development
falls with reserve size, and transfer each movement to a target's size using the fitted scales and a tail map where
regimes differ. Every transferred
scenario can be traced to a named filing, so the library behind a stress can be inspected.

\pnum Size improves held-out density prediction materially. But the transferred extreme tail understates the smallest
books: at 99.5\% three syndicates supply the six misses, and each discloses exposure still earning or old years still
carrying exposure, whose contribution is unknown.
Plain square-root scaling has descriptively lower mean pinball loss at 99\% and 99.5\%. So the transfer makes the size adjustment explicit and supports scenario comparison; it does not calibrate a 1-in-200.

\subsection{What it means in practice}
\label{sec:conclusion:practice}
\pnum For each proposed donor, inspect the claims measure, the remaining exposure, the gross reserve base and the
business that generated the liabilities. Then show the transfer and the donors behind the resulting tail. A firm should
back any diversification credit or load with evidence about its own portfolio, not adopt the illustrative
transferred quantile.

\paragraph{Acknowledgements}
Generative AI was used in this work in four distinct roles, all declared here in
full.

\emph{As a method, within the data pipeline.} Structured data were extracted from
the syndicate reports using a dual-model design: Google \texttt{gemini-2.5-flash} and OpenAI
\texttt{gpt-5-mini}, accessed through their respective APIs, independently read report text.
Not every report has two usable model outputs. Where both models supply a comparable field, their readings are compared; deterministic table
extraction is used where available under the precedence rules below. An absolute-amount triangle ordinarily takes precedence, subject to the
provisions sign-disagreement override and the model-agreement veto stated in the
data section. A managed- or group-level loss-ratio triangle instead fills a blank narrative
value, or overrides a syndicate-specific narrative value only when their directions
contradict. The two readings disagree on a numeric field where they differ by more than
$0.5\%$ of the larger and by more than $0.05$ in absolute terms; text, list, page and
confidence fields are exempt. A disagreement is logged but
excludes no record, and the unresolved residue is reported in \ref{app:audit:caveats}. Where a rule in this
pipeline was later corrected, the records re-extracted under it are listed
individually, with the development figure and adoption route on both sides, in the
extraction repository's generated change log; the counts reported here are after
those corrections. The presentational-currency scan is deterministic: an explicit
presentation statement, else the dominant monetary unit-header, else a
functional-currency statement, each recorded with an evidence quote and page reference
for audit. The two models' currency field decides only the filings with no usable text
layer, 256 of the 1{,}065 reports. The
external-RITC scan is a deterministic sentence classifier, not a language model,
and records the sentence, its direction and effective year, and the page.
The dates and counts that follow are read from the committed records by the build.
The committed response caches date every model call they hold: 1{,}772 on 16--20 March, 850 on 5--6 July, 4 on 30 August, 49 on 5 September, 3{,}117 on 11 September, 1 on 13 September, 32 on 29 September and 12 on 3 October 2026 (extraction commit \texttt{c9a6f75}).
The run log records extraction runs from 12 March to 3 October 2026, under prompt versions 1.8 to 2.13; its runs of 12--15 March precede the earliest retained cache, and its runs of 9, 10, 12 and 21 September cached no response.
Of those responses, 77 are from a third model, Anthropic's \texttt{claude-sonnet-4-6}, the pipeline's adjudicator, accessed through its API: 76 on 16--20 March and 1 on 11 September 2026, on opening reserves (26), development (21), direction (16), currency (8) and other fields (6).
In the pipeline's interactive mode, where the two extraction models disagree beyond
tolerance, the adjudicator reads the filing and proposes a value, which the author
accepts or overrides. The proposals and the decisions are recorded in the extraction
repository's disagreement and rejection logs. No value or exclusion in the committed
records comes from them. The analysis reads the disagreement log only to publish its
pending, resolved and prompt-version workflow counts; it does not use the log to select
a value or exclude a record.
The prompt-version log, per-field provenance and disagreement
logs are in the extraction repository cited in the Data Availability Statement. This use is a component of the method and is described in
Section~\ref{sec:data} and \ref{app:audit}, not merely disclosed here.

\emph{As a method, in measuring the extraction error rate.} The records of the
error-rate study (\ref{app:audit:errorrate}) were read against their filings by
Anthropic's Claude Opus 5 (\texttt{claude-opus-5}), through the Claude Code
application. The first readings were made by Claude Code agents, each given the
protocol's definitions, a brief of the figure the loader adopted, an evidence pack and
a tool that prints the filing's pages. The second readings were made by the Claude
Code session that ran the study, except for the records read from 21 September 2026 on,
which Claude Code agents read two to a record, independently of each other. No person
read the filings for the study, and every
reading is published with the page and the arithmetic it rests on.
The study's samples were drawn and its records read on 13--14 September 2026, 21--22 September 2026 and 13 October 2026; its files are at analysis commit \texttt{6b668fc}, in \texttt{results/extraction\_error\_rate/}.

\emph{As a drafting and coding aid.} Anthropic's Claude was used through the Claude
Code application to write and revise the code, tests and documentation of the
extraction and analysis repositories, to convert the author's draft text into this
manuscript's \LaTeX{} source, begun on 25 June 2026, and to edit and copy-edit it.
Each commit Claude Code made names its model, and the models and dates here are the
ones the three repositories' commits record.
The commits name Claude Opus 4.6 (27 February to 3 April 2026: the extraction and analysis repositories), Claude Opus 4.8 (25 June to 12 July 2026: the analysis repository and this manuscript's source), Claude Fable 5 (6 July to 3 September 2026: the extraction and analysis repositories and this manuscript's source), Claude Opus 5 (29 August to 26 September 2026: the extraction and analysis repositories and this manuscript's source), Claude Fable 5.1 (3 to 7 September 2026: the extraction and analysis repositories and this manuscript's source), Claude Opus 5.5 (29 September to 1 November 2026: the extraction and analysis repositories and this manuscript's source) and Claude Sonnet 5.5 (30 September to 28 October 2026: the extraction and analysis repositories and this manuscript's source). Claude Code's helper agents (the implementer, verifier and reviewer) ran on Claude Sonnet 5.5 from 2 to 8 October 2026, and the commits they wrote in those days carry the session's Claude Opus 5.5 trailer, so the trailers understate Sonnet 5.5's use (the author's statement). Claude Code's adversarial paper reviewer agent, which ran on Claude Sonnet, read the manuscript and code from 1 September 2026 and reported errors; it made no edits (the author's statement).
Claude Opus 4.8 was also used, through the Claude application, on earlier versions
of this manuscript, which predate the \LaTeX{} source; that is the author's
statement, as no commit records it.

\emph{As a checking and editing aid.} OpenAI Codex was used to check the manuscript,
including its statistical claims and their agreement with the analysis outputs.
GPT-5.6 Sol was used for earlier checking; GPT-6 was used in October 2026 to review
source accounts, correct maturity-proxy data, run sensitivity analyses, and edit the
manuscript and supporting code.

No AI system is an author. The research question, model specification, data
construction decisions, statistical analysis, choice of robustness checks, results
and their interpretation are the author's own. AI output was reviewed and, where
used, verified by the author, except the error-rate study's readings, which are
published so that any reader can check them. The author is entirely responsible for the scientific
content of this paper.

\paragraph{Funding}
This research received no specific grant from any funding agency, commercial or
not-for-profit sectors.

\paragraph{Competing Interests}
None.

\paragraph{Data Availability Statement}
The reserve-movement data underlying this article are derived from publicly
available Lloyd's syndicate reports and accounts for reporting years 2014--2024
and can be reconstructed from those sources. The source reports are not redistributed.
The code repositories and the cited commits are public:

\begin{quote}\raggedright\small
report download and structured-data extraction pipeline (commit \texttt{c9a6f75}):\\
\url{https://github.com/colinpriest/lloyds-reserve-stress-testing}\\[4pt]
analysis code reproducing the results (commit \texttt{6b668fc}):\\
\url{https://github.com/colinpriest/lloyds-exposure-composition-analysis}
\end{quote}

\noindent All empirical results and the analysis-generated figures and tables are produced from those commits. The
text of this article and of the online supplementary material (\hbox{Supplement~S1}--\hbox{S24}),
Figure~\ref{fig:workflow}, the expository tables (the measured movement, the tail cases, the uses, notation, priors and
model alternatives) and the filing evidence the supplement quotes are authored in the manuscript source; they quote
numbers from those commits and words from the filings. The analysis uses the committed extracted records; its full
manifest also reads estimator code from the pinned extraction commit in a separate Git checkout and refreshes
Federal Reserve H.10 exchange rates over the network. Neither the source reports nor an application programming
interface (API) key is needed for that analysis. Re-running the extraction needs the source reports: fresh model and optical character recognition
inference also needs paid API access, while cached replay re-derives the extraction from the committed response caches
without it.
The last replay of the deterministic step could not serve 14 records, whose committed values stand;
re-deriving them would need fresh paid inference. The replays' measured coverage, the reports they cannot serve and
the analysis's run report are recorded in \supp{12}; the last replays were not made at the cited extraction commit,
\texttt{c9a6f75}.
The extraction repository
also holds the machine-readable coverage audit (\texttt{coverage\_status.xlsx}).

\printbibliography

\appendix

\clearpage
\section{Notation and estimation detail}
\label{app:notation}

\subsection{Notation}
\label{app:notation:symbols}
Table~\ref{tab:notation} collects the symbols used in the body; each is also
defined in prose at first use. Two scales are kept apart. The base scale $s(R)$
depends on size alone and is the one a transfer uses. The fitted observation scale
$\sigma_{i,t}$ multiplies it by the reporting-year shock and the RITC scale term, and
is the one the likelihood uses.

\begin{table}[H]
\begin{threeparttable}
\caption{Notation used in the body.}
\label{tab:notation}
\begin{tabular}{lp{0.66\linewidth}}
\toprule
\headrow Symbol & Meaning \\
\midrule
$(i,t)$ & syndicate $i$ in reporting year $t$ (the unit of analysis) \\
$N$, $J$ & records (632) and syndicates (117) in the primary sample \\
$R_{i,t}$ & reserve base and size measure: opening gross claims outstanding, plus gross reserves taken on in the
year whose development $M_{i,t}$ includes (Section~\ref{sec:data:source}) \\
$M_{i,t}$ & signed one-year change in gross claims estimates on the underwriting years $u\le t-2$ (Section~\ref{sec:data:pyd}) \\
$S_{i,t}=M_{i,t}/R_{i,t}$ & the measured movement, or severity (adverse positive) \\
$R_{\mathrm{ref}}$ & reference size, 500m \\
$s(R)$ & base size scale, $\sqrt{\sigma_{\mathrm{undiv}}^{2}+\sigma_{\mathrm{div}}^{2}(R/R_{\mathrm{ref}})^{2(k-1)}}$ \\
$\sigma_{i,t}$ & fitted observation scale, $\exp(\eta_t+\beta_{\mathrm{RITC}}\mathbf 1_{\mathrm{RITC}})\,s(R_{i,t})$ \\
$k$ & pooling exponent, $k=\tfrac12+\tfrac12\operatorname{logit}^{-1}\theta$, so confined to $(\tfrac12,1)$ by the prior \\
$\sigma_{\mathrm{undiv}}$ & undiversifiable floor (a Student-$t$ scale) \\
$\sigma_{\mathrm{div}}$ & diversifiable Student-$t$ scale at $R_{\mathrm{ref}}$ \\
$\nu_{\mathrm{clean}},\ \nu_{\mathrm{RITC}}$ & clean- and RITC-regime degrees of freedom \\
$\lambda_{\mathrm{RITC}}$ & log tail-weight shift, $\nu_{\mathrm{RITC}}=\nu_{\mathrm{clean}}e^{-\lambda_{\mathrm{RITC}}}$ \\
$\beta_{\mathrm{RITC}}$ & RITC log-scale term (in $\sigma_{i,t}$, not in a transfer) \\
$\eta_t\sim\mathcal N(0,\tau)$ & reporting-year shared shock; $\tau$ its log-scale standard deviation \\
$c_i$ & transfer factor, target base scale over donor base scale, $s(R^{(q)})/s(R_{i,t})$ \\
$F_\nu$ & Student-$t$ CDF (the RITC rank map, Equation~\eqref{eq:operator}) \\
\midrule
\multicolumn{2}{l}{\emph{Concentration overlay only (Section~\ref{sec:framework:concentration})}} \\
$w_{i,t,\ell}$, $H_{i,t}$ & premium share of line $\ell$ (13-category taxonomy); $H_{i,t}=\sum_\ell w_{i,t,\ell}^{2}$ \\
$\gamma$, $R^{\mathrm{eff}}_{i,t}$ & concentration exponent (0 in the headline); effective size $R_{i,t}(1/H_{i,t})^{\gamma}$, which replaces $R$ in $s(\cdot)$ \\
\bottomrule
\end{tabular}
\end{threeparttable}
\end{table}

\subsection{Sampling and convergence}
\label{app:notation:sampling}
The model was fitted in PyMC~5.25.1 (PyTensor~2.31.7, NumPy~2.1.3, SciPy~1.14.1;
Python~3.12.6) by the No-U-Turn Sampler \autocite{HoffmanGelman2014}, running four
chains of $1{,}500$ tuning and
$1{,}500$ post-warmup draws ($6{,}000$ posterior draws) at
\texttt{target\_accept}\,$=0.98$ under random seed~42. Convergence was clean: the
maximum split-$\hat R$ was $1.00$, there were no divergent transitions, and the minimum bulk effective sample size was $1{,}416$. Auxiliary analyses use
their own fixed seeds, recorded in the result files. Models are
compared by held-out predictive density, from five-fold cross-validation by syndicate or
from Pareto-smoothed importance-sampling leave-one-out cross-validation (PSIS-LOO)
\autocite{Vehtari2017}, as \ref{app:robust} describes.

\subsection{Prior specification}
\label{app:notation:priors}
The estimated parameters carry weakly-informative priors (Table~\ref{tab:priors}). The floor
and diversifiable scales are deterministic functions of a total-dispersion scale and a
Beta-distributed variance share, so within the floored specification the share is estimated,
not fixed; that does not identify the existence or the extrapolated level of a floor
(\ref{app:robust:floor}). The pooling exponent $k$ is the logistic transform of $\theta$, and the
location is fixed at $\mu=0$ (\supp{16}).

\begin{table}[H]
\begin{threeparttable}
\caption{Prior distributions for the estimated parameters.}
\label{tab:priors}
\begin{tabular}{lll}
\toprule
\headrow Parameter & Support / transform & Prior \\
\midrule
$\theta$, with $k=\tfrac12+\tfrac12\operatorname{logit}^{-1}\theta$ & unconstrained & $\mathcal N(0,\,1.5)$ \\
$\gamma$ ($\gamma$-free fit; 0 in the headline) & $\ge0$        & $\text{HalfNormal}(1.0)$ \\
$\log\sigma_{\mathrm{tot}}$     & unconstrained & $\mathcal N(\log 0.05,\,1.0)$ \\
$f$ (undiv.\ variance share)    & $[0,1]$       & $\text{Beta}(1,1)$ \\
$\nu_{\mathrm{clean}}$          & $>0$          & $\text{Gamma}(2,\,0.1)$ \\
$\tau$                          & $\ge0$        & $\text{HalfNormal}(0.5)$ \\
$\lambda_{\mathrm{RITC}}$       & unconstrained & $\mathcal N(0,\,0.7)$ \\
$\beta_{\mathrm{RITC}}$         & unconstrained & $\mathcal N(0,\,0.5)$ \\
\bottomrule
\end{tabular}
\begin{tablenotes}[hang]
\item[] Normal and HalfNormal are parameterised by their standard deviation;
$\text{Gamma}(\alpha,\beta)$ by shape $\alpha$ and rate $\beta$ (prior mean
$\alpha/\beta=20$ for $\nu_{\mathrm{clean}}$, so heavy tails are allowed but not
forced). The scales are deterministic,
$\sigma_{\mathrm{undiv}}=\sigma_{\mathrm{tot}}\sqrt{f}$ and
$\sigma_{\mathrm{div}}=\sigma_{\mathrm{tot}}\sqrt{1-f}$ with
$\sigma_{\mathrm{tot}}=e^{\log\sigma_{\mathrm{tot}}}$; the $\text{Beta}(1,1)$ share
leaves the variance-share posterior unshrunk toward zero within the floored
specification, which is not the same as identifying the floor
(\ref{app:robust:floor}). Year effects are
non-centred: $z_t\sim\mathcal N(0,1)$ and $\eta_t=\tau z_t$.
\end{tablenotes}
\end{threeparttable}
\end{table}

\subsection{Remaining posterior summaries}
\label{app:notation:nuisance}
Table~\ref{tab:nuisance} gives the rest of the headline posterior beside Table~\ref{tab:posterior},
and the concentration exponent of the $\gamma$-free fit that supplies the optional overlay. The RITC
terms are weakly informed: the posterior gives $P=0.40$ that the RITC tail is the heavier, against
$0.50$ under the prior. The overlay's $\gamma$ is barely informed by the data: $P(\gamma>0.05)=0.93$,
against $0.96$ under its prior.

\begin{table}[H]
\begin{threeparttable}
\caption{Remaining posterior summaries: the headline fit ($\gamma$ fixed at $0$; $n=632$) and the
concentration exponent of the $\gamma$-free fit.}
\label{tab:nuisance}
\begin{tabular}{llcc}
\toprule
\headrow Parameter & Meaning & Mean & 95\% HDI \\
\midrule
$\nu_{\mathrm{RITC}}$     & RITC-regime degrees of freedom (derived)  & 5.39  & [1.13, 11.80] \\
$\lambda_{\mathrm{RITC}}$ & log tail-weight shift on RITC records     & \textminus0.166 & [\textminus1.225, 0.885] \\
$\beta_{\mathrm{RITC}}$   & RITC log-scale term                        & \textminus0.095 & [\textminus0.487, 0.271] \\
$\tau$                    & reporting-year shock standard deviation (log) & 0.137 & [0.000, 0.267] \\
\midrule
$\gamma$ ($\gamma$-free fit) & concentration exponent, the overlay's source & 0.394 & [0.000, 0.932] \\
\bottomrule
\end{tabular}
\begin{tablenotes}[hang]
\item[] Posterior probabilities, each beside the value the prior alone gives:
$P(\sigma_{\mathrm{undiv}}>0.005)=0.995$ (prior $0.94$); no draw of $\nu_{\mathrm{clean}}$ below $2$ in
$6{,}000$ (prior $0.018$); $P(|\beta_{\mathrm{RITC}}|>0.1)=0.64$ (prior $0.84$); $P(\nu_{\mathrm{RITC}}<2)=0.05$
(prior $0.04$). Both $\sigma$ parameters are Student-$t$ scales, not standard
deviations. Intervals are highest-density intervals (HDIs); $R_{\mathrm{ref}}=500$m; $\mu=0$ is fixed;
monetary figures are sterling-converted (\supp{11}). The $\gamma$-free fit is compared with the
headline in \supp{19}.
\end{tablenotes}
\end{threeparttable}
\end{table}

\section{Data audit}
\label{app:audit}
This appendix reconciles the filings to the sample and describes the extraction, its error rate and
unresolved residue. The supplement covers extraction (\supp{12}), selection (\supp{6}), reporting
years and run-off (\supp{7}), business mix (\supp{8}), and reinsurance to close (\supp{9}).

\subsection{Coverage and reconciliation}
\label{app:audit:exclusions}

\begin{table}[H]
\begin{threeparttable}
\caption{Reconciliation from the retrieved filings to the working sample,
2014--2024.}
\label{tab:reconcile}
\begin{tabular}{lrr}
\toprule
\headrow Step & Records & Running total \\
\midrule
Filings retrieved & -- & 1{,}065 \\
\quad less: \parbox[t]{0.45\linewidth}{\raggedright no deterministic reading (the parsers found no prior-year figure; the models were not run)\strut} & \textminus20 & 1{,}045 \\
\quad less: \parbox[t]{0.45\linewidth}{\raggedright loader skips (disclosure screen, unresolved eligibility or unavailable eligible outcome)\strut} & \textminus107 & 938 \\
\quad less: \parbox[t]{0.45\linewidth}{\raggedright no usable development reading (the extraction's validation failed and no model carries a development percentage)\strut} & \textminus4 & 934 \\
\quad less: \parbox[t]{0.45\linewidth}{\raggedright in run-off (the filing states run-off for the whole year, with or without a development figure, outside the RITC regime; or gross written premium $=0$, or $<0$ where the filing states run-off)\strut} & \textminus28 & 906 \\
\quad less: \parbox[t]{0.45\linewidth}{\raggedright no reserves above \pounds0.1m (none read, nil, or positive at or below the floor)\strut} & \textminus4 & 902 \\
Corpus & -- & 902 \\
\quad less: development on a net or unstated basis & \textminus98 & 804 \\
\quad less: take-on, not development & \textminus8 & 796 \\
\quad less: unusable severity & \textminus93 & 703 \\
\quad less: missing opening reserves & 0 & 703 \\
\quad less: missing line-of-business weights & \textminus71 & 632 \\
\textbf{Working sample} & -- & \textbf{632} \\
\bottomrule
\end{tabular}
\begin{tablenotes}[hang]
\item[] The extracted syndicate-years carry no duplicate (syndicate, year) pairs.
Each filing appears in exactly one row: the steps are the loader's sequential
dispositions (a record-level ledger is in the
analysis repository), so the running totals are exact.
The unusable-severity step includes the 11 records the source-history rule sets aside
(Section~\ref{sec:data:filters}); with them back, the sample is the 643-record comparator.
The running total of $1{,}045$ after the first step equals the count of active syndicate-years, but the
two are different sets: the active count comes from Lloyd's lists and the filings from the retrieval, and
corpus records off their year's list belong to the second only (\supp{7}).
The first step's filings carry the extraction status no deterministic reading, and may still
print a development table the parsers could not read.
In the second step, the extraction found no prior-year figure in the reserve text of 60 of the
107 filings and did not read that of 10. Another 24 had no deterministic reading, and their
filings state that the syndicate began underwriting in the report year or the year before. Recent inception alone does not rule out inherited older liabilities. Another 11 state a
prior-year figure, but their own triangles hold no underwriting year up to $t-2$. The first-year report 6118/2014 participated by a quota share in older underwriting years; its opening claims reserves and mature-cohort movement remain unresolved. The remaining case, 1254/2022, is an eligible mature-cohort RITC take-on whose gross
development cannot be reconciled to its accounts and is therefore unavailable.
The 71 records without line-of-business weights are the 37 whose extracted mix names no
line and the 11 life records, out of scope (\supp{8}), and the 23
composition-unavailable records.
Separately, 117 of the 1{,}065
retrieved filings carried no usable dual-model extraction; that extraction diagnostic
overlaps the first two steps and is not an inferential missing-outcome
population or a step in this flow. The basis exclusion is
described in Section~\ref{sec:data:filters} and \supp{12}. The working
sample (632) and the active syndicate-years (1{,}045) are different populations, so their ratio is not a
coverage. Of the 473 syndicate-years on Lloyd's official lists for 2020--2024, 310 are in the working sample
(66\%; \supp{7}).
Syndicate-years in the RITC regime (an accepted RITC or a confirmed inward
transfer) are retained and modelled as a separate tail regime, not excluded; a record whose filing
shows its adopted figure to be the take-on itself is not development and leaves the
working sample (\supp{9}).
\end{tablenotes}
\end{threeparttable}
\end{table}

Table~\ref{tab:reconcile} traces the retrieved filings to the working sample. A filing enters
the \emph{corpus} once it has a usable development reading, its opening reserves are above
\pounds0.1m and it is not a run-off year, whatever the basis of its development figure. It enters
the \emph{working sample} once the figure is on the gross basis and its severity, opening reserves
and line-of-business weights are all complete; the run-off rules are in \supp{7}. One working-sample record,
1840/2022, is a zero movement: its only mature cohort's estimates, paid and reserve entries are printed as dashes,
which we read as nil (\supp{6}). Coverage is
uneven by year. 2014 contributes no record: its reports predate FRS~103, and most of its corpus
records state development net or on an unstated basis. For 2020--2024 the working sample holds from 59\% to 76\% of each year's listed syndicates. No share is given for 2015--2019: the sources assembled for this study give counts but no individual membership lists for those years, so membership cannot be checked from them (\supp{7}).

Selection runs by size (Section~\ref{sec:data:filters}). A provisional disclosure-based target holds 666 records: the 632 of the
working sample, 24 whose premium mix is unavailable and 10 treated as eligible with unavailable outcomes.
The 24 are the 23 composition-unavailable records of Table~\ref{tab:reconcile} and 1 that leaves earlier, at the
unusable-severity step.
For nine of those ten, the applicable mature outcome and reserve base remain uncertain (\supp{6}).
Weighting the likelihood by a size-and-year response propensity moves $k$ from $0.624$ to $0.672$; the
pseudo-outcome stresses also depend on the assumed target and the placement of synthetic outcomes (\supp{6}). None of these exercises
establishes an ignorable selection mechanism, so missing-at-random cannot be established.

\subsection{Provenance and precedence}
\label{app:audit:provenance}
The extraction searches report pages for claims-development triangles, provisions notes,
segmental premium tables and balance sheets. The records preserve extraction routes and source
references where available; a search does not guarantee that a usable disclosure is found. The
movement $M_{i,t}$ is ordinarily computed from the printed gross absolute-amount triangle over the underwriting
years $u\le t-2$, with the source-evidenced corrections described in Section~\ref{sec:data:pyd}. In the extraction it is subject to two
rules: where the triangle and the provisions movement disagree
in sign, the provisions figure is preferred, and where both models agree on the opposite sign, or the
figure exceeds $50\%$ of the opening reserves while both model movements are below $10\%$, the models'
reading stands. Where no absolute-amount triangle gives a figure, a loss-ratio triangle provides a
conditional fallback. The primary sample selects figures intended to measure mature-cohort development;
3330/2017 uses a roll-forward estimate whose cohort scope is uncertain (\supp{2}). The source rule sets a figure aside
only on evidence about it, never on sign alone. The full method, its screens and its adjudications are
in \supp{12}.

\subsection{Extraction error rate}
\label{app:audit:errorrate}
The error rate is the probability that a record drawn at random from the working sample has a
materially wrong adopted development figure. A figure is materially wrong when it differs from the
filing's figure by more than the larger of 0.5m and $5\%$ of the adopted figure, in the report's
currency; when its sign is opposite; or when its scope is plainly wrong (it takes in the two most
recent underwriting years, is a cohort's cumulative total, or is net where the record claims gross).
A triangle figure is checked against the triangle recomputed from the filing's printed table.
Records the filing cannot settle leave the denominator. The rate measures fidelity to the paper's
numerator, not the economic basis of the figure. A second $5\%$ line applies to the stress: a change
of more than $5\%$ in the transferred $99.5\%$ quantile of a worked target is material, and the
simulation of errors not yet found reads its effect against that line (\supp{12}).

The rate was measured on random draws, each read against the filing page by page under a protocol
fixed before the draw. Earlier draws and censuses of single mechanisms found errors, and each
mechanism was repaired before the final fit; that history is in \supp{12}. The fourth draw, on 13 October 2026,
measured the extraction of that date. Its frame was the 400 records of the then 674-record working sample that no draw
or census had read. Its 100 records held 2 errors, 96 correct figures and 2 undeterminable ones: a posterior mean of
$2.5\%$ under a Jeffreys prior, with a $95\%$ credible interval from $0.4\%$ to $6.4\%$. Both errors read a gross
table's net companion in its place, and a census then repaired or excluded every triangle-sourced case of that
mechanism (\supp{12}). The draw's 94 records still in the primary sample are reconciled with the figures it now adopts.
Thirteen of their figures have changed, each is scored only on a reading of it, and none of the 94 now carries a known
error. The readings that score them include those the repairs were taken from, so this shows that the figures now
adopted are the ones the readers read; it is not an independent check. That is not a rate: the 94 remain from an earlier population, and the draw's errors were repaired because it
found them. For the 247 of the 632 records classified as unread by the simulation, it assumes the historical $2.5\%$
error rate (\supp{12}), an assumption, not a measurement. Two other records are treated as read on the basis of source
readings supporting their derived figures, although neither final figure was directly read.

\subsection{Unresolved extraction residue}
\label{app:audit:caveats}
The two models' readings of the report text disagree in places. The extraction repository's
disagreement log holds 542 cross-model discrepancies, each a report and a field, pending review,
201 of them from prompt version 2.13, the version of the latest extraction run. A discrepancy
excludes no record: the loader takes one model's reading by a fixed rule (\supp{12}). 57 of the 632
working-sample records carry such a disagreement. The error-rate study above measured what such
residues left in the adopted figure when it was drawn, on 13 October 2026. The other residual limits of the extraction are in \supp{12}.

\section{Model comparison and robustness}
\label{app:robust}
This appendix holds the comparisons behind the size law: the pooling exponent scored
against its two benchmarks, the floor scored against a floorless law, the goodness-of-fit
diagnostics, and a table of every model choice with the evidence for it. Further
robustness and sensitivity work is in \supp{13}--\supp{20}.

The scale-law comparisons are scored by held-out log predictive density on five folds of
syndicates, refitting every model in each fold; the free-exponent model with a floor had $1$
divergent transition in one fold, the no-floor model $3$ in one fold, and every other fold fit
had none. Each comparison is summarised by the posterior of the score difference under a Bayesian
bootstrap over syndicates (Dirichlet weights on the syndicate totals of the pointwise
differences). The RITC scale term (\supp{15}) and the reserve-cycle extensions (\supp{18}) are
scored by observation-level PSIS-LOO instead, a weaker check because a syndicate's other years
stay in the fit. The few \emph{model-agnostic} diagnostics (rank correlations, permutation tests,
Kruskal--Wallis and Mann--Whitney comparisons) check the model without assuming it; no inference
rests on them, and a printed $p$ describes how far the sample sits from that diagnostic's null.

\subsection{Pooling: the free exponent against its two benchmarks}
\label{app:robust:pooling}
The free exponent $k\in(\tfrac12,1)$ is scored against its two benchmarks on the same five
syndicate folds (Table~\ref{tab:poolingcv}): $k=\tfrac12$, the finite-variance independent-pooling benchmark for the diversifiable term, and
$k=1$, a constant scale. All three carry the floor. They are single-regime cross-validation fits
(one Student-$t$ tail, no reporting-year shock) that carry the concentration term with $\gamma$
estimated, as the $\gamma$-free fit does.

\begin{table}[H]
\begin{threeparttable}
\caption{Pooling exponent under syndicate prediction: the free exponent against $k=\tfrac12$ and
against the constant scale $k=1$, all carrying the floor, on the five syndicate folds of
Section~\ref{sec:results:oos}.}
\label{tab:poolingcv}
\begin{tabular}{lcccl}
\toprule
\headrow Exponent setting & Held-out ELPD & Free minus setting & $P(\text{free }k\text{ better})$ & Reading \\
\midrule
Free $k\in(\tfrac12,1)$ & 701.1 & --                   & --   & reference \\
$k=\tfrac12$ (diversifiable term) & 699.8 & +1.2 [\textminus4.0, +6.2] & 0.70 & not separated \\
$k=1$ (constant scale)            & 667.8 & +33.3 [+9, +63] & 0.9972 & strongly disfavoured \\
\bottomrule
\end{tabular}
\begin{tablenotes}[hang]
\item[] $n=632$, 117 syndicates. Differences are calculated before rounding the scores, so the displayed
difference can differ from the difference of the displayed scores. A positive difference favours the free exponent; intervals are $95\%$
equal-tailed credible intervals and probabilities are posterior probabilities, both under the Bayesian bootstrap over
syndicates. The $k=1$ row is the naive market pool, fitted separately, so its score lies $0.1$ from the naive
pool's in Table~\ref{tab:oos}. The headline's size-only form leads the naive pool by $+34.4$ $[+11,+64]$,
$P=0.9979$ (Table~\ref{tab:oos}). The observation-level PSIS-LOO comparison of the two lower settings, with its
stacking weights, is in \supp{13}.
\end{tablenotes}
\end{threeparttable}
\end{table}

The constant-scale specification ($k=1$) is strongly disfavoured by held-out density
prediction. Fixing $k=1$ removes the size channel, so the model is the naive market pool of
Section~\ref{sec:results:oos}. This comparison, not the posterior's $P(k<1)$, which is one by
construction inside the bracket, is the evidence for size dependence. A marginal scale law does not
determine the joint distribution of movements across syndicates, so it is not a test of every
comonotonic joint law.

The lower benchmark is not separated: against $k=\tfrac12$ the free exponent scores $+1.24$
$[-4.0,+6.2]$, $P=0.70$, and is ahead on $44\%$ of held-out syndicate-years. In these comparison fits
(single-regime, with $\gamma$ free), the fixed-half model compensates with a larger floor ($\sigma_{\mathrm{undiv}}$ from $0.030$ to $0.039$): a reallocation
within the fitted marginal scale law, not an identified dependence component. Refitting the adopted
two-regime model with $k$ fixed at $\tfrac12$ moves the worked target's transferred $99.5\%$ quantile
under the size-only transfer from $0.342$ to $0.315$, and the $100$m/$2{,}000$m scale ratio from $2.19$
to $2.07$ (\supp{13}). So the choice between the free exponent and $k=\tfrac12$ is unresolved, and it
moves the stress by about $8\%$.

The bracket is a prior, so its lower end was also released, to $(0.25,1)$ and to $(0,1)$. The
posterior means are $0.610$ and $0.609$, with $P(k<\tfrac12)=0.07$ under both, against $0.32$ and
$0.50$ under the two priors, and no pile-up of draws at an edge of either support (\supp{13}). So the
data place most of the exponent's mass above $\tfrac12$ without the bracket's help, and do not
exclude a value below it. The worked target's transferred $99.5\%$ quantile falls by $1.5\%$ in both
fits.

\subsection{Predictive support for the undiversifiable floor}
\label{app:robust:floor}
A floor bounded away from zero within the floored model's posterior is not a model comparison,
and the pooling comparison above holds a floor in \emph{both} candidates. So the floorless
alternative is scored directly, on the same by-syndicate folds (Table~\ref{tab:floorstrata}). Both
forms are the cross-validation fits; the headline's own floor is not scored out of sample.

\begin{table}[H]
\begin{threeparttable}
\caption{Floor versus no floor, decomposed by reserve size. $\Delta$ is held-out
ELPD for the floor model minus the no-floor model on the same by-syndicate folds,
so a positive value favours the floor.}
\label{tab:floorstrata}
\begin{tabular}{lrrrcc}
\toprule
\headrow Reserve size & Obs. & Syndicates & $\Delta$ELPD & 95\% credible & $P(\text{floor better})$ \\
\midrule
All                       & 632 & 117 & +1.35 & [\textminus2.59, +5.48] & 0.75 \\
\midrule
Below 100m              & 90 &  48 & \textminus0.13 & [\textminus3.03, +3.40] & 0.43 \\
100m--500m            & 269 &  70 & +0.94 & [\textminus0.60, +2.37] & 0.89 \\
500m--2bn             & 222 &  49 & \textminus0.53 & [\textminus1.49, +0.38] & 0.13 \\
Above 2bn               &  51 &  10 & +1.07 & [\textminus1.06, +2.72] & 0.87 \\
\bottomrule
\end{tabular}
\begin{tablenotes}[hang]
\item[] Bayesian bootstrap with Dirichlet weights on the syndicate totals within
each stratum, so repeated years from one syndicate are not treated as independent;
each probability is a posterior probability under that bootstrap, the share of $20{,}000$ draws on which the floor model scores better. No working-sample syndicate-year exceeds $5$bn (the largest
reserve base is $4.75$bn), so no stratum is reported there.
\end{tablenotes}
\end{threeparttable}
\end{table}

Pooled over all $632$ syndicate-years the floored model scores slightly better on held-out
data, $\Delta\text{ELPD}=+1.35$ $[-2.59,+5.48]$, $P(\text{floor better})=0.75$, which does not
separate the two, and no stratum separates them either. The floorless model has the larger exponent,
$k=0.745$ $[0.685,0.803]$ against $0.632$ $[0.522,0.742]$ (both single-regime, without the
reporting-year shock): the exponent--floor trade-off that the posterior correlation of $-0.71$ records
(\ref{app:robust:ident}). One floorless fold fit had divergent transitions, which can lower its held-out
score, so the comparison may be tilted toward the floor by an amount not measured.

The two fitted scale curves lie within about $4\%$ of each other between $50$m and $1$bn, where
$451$ of the $632$ observations sit, and move apart above about $2$bn, where the stratum holds 51
syndicate-years from 10 syndicates; at the largest observed book, $4.75$bn, the floorless scale is
about $17\%$ below the floored one (\supp{14}). The conditional size trend above $1$bn is equally
unresolved: $-0.004$ (95\% HDI $[-0.42,+0.40]$), against the $-0.255$ a floorless law implies
(\supp{14}). So the two specifications describe the observed range almost equally well and
differ in what they imply beyond it, where a positive asymptote is retained as a prudential
extrapolation (Section~\ref{sec:results:posterior}). Settling the floor would need more very large
books, or another source, such as regulatory returns, that resolves reserve volatility at
multi-billion scale.

\subsection{Identification and serial-dependence checks}
\label{app:robust:ident}
In the headline posterior $k$ trades off with the floor (correlation $-0.71$) and the diversifiable
scale ($+0.69$); the design is not collinear (\supp{14}, \supp{17}).

\raggedbottom
\subsection{Goodness of fit and calibration}
\label{app:robust:fit}
\begin{figure}[H]
\centering
\includegraphics[width=0.9\linewidth]{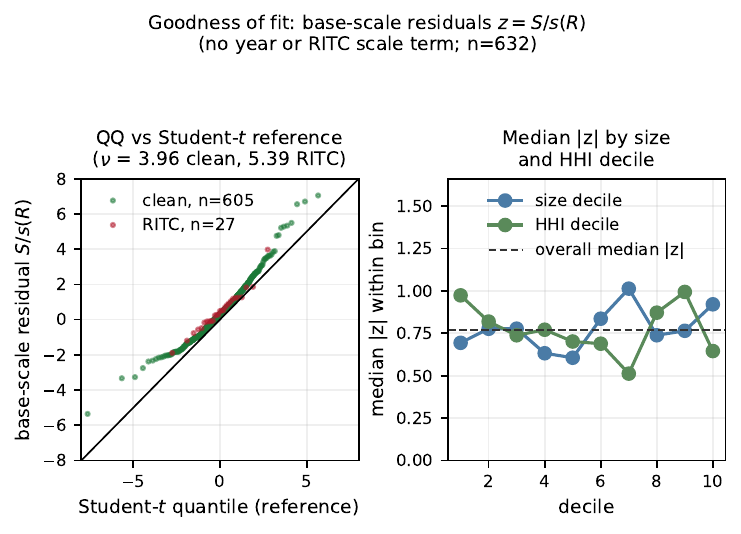}
\caption{Goodness of fit with the base-scale diagnostic $z=S/s(R)$, at the posterior means,
without the reporting-year shock or the RITC scale term. Left: quantile--quantile (QQ) plot of $z$ for
each regime against a Student-$t$ reference (clean $\nu=3.96$, RITC $\nu=5.39$); the reference is
not the full model's conditional distribution of $z$. The clean points lean adverse in the upper
tail; the RITC points lie on a flatter line against their reference. The base-scale residual retains the realised
year and RITC scale effects. Right: median
$|z|$ in each size and concentration decile, with the overall median dashed: $0.61$ to $1.01$
across the size deciles and $0.51$ to $0.99$ across the concentration deciles, with no monotone
trend along either axis. Similar central values do not show that the shapes or tails agree.}
\label{fig:gof}
\end{figure}

The fitted scale law is checked through a base-scale diagnostic,
$z_{i,t}=S_{i,t}/s(R_{i,t})$, read at the posterior-mean parameters. It divides by the base size
scale alone. It leaves out the reporting-year shock and the RITC scale term of $\sigma_{i,t}$, so
it is not the full model's conditional residual. If the pooling law is right, $z$ should show no
trend in level along the size axis. Across the ten reserve-size groups the median of $|z|$ runs from
$0.61$ to $1.01$ with no monotone trend (Figure~\ref{fig:gof}), and a Kruskal--Wallis rank test, a
model-agnostic description that is not calibrated here, gives $p=0.21$. Across the concentration
deciles the medians run from $0.51$ to $0.99$ ($p=0.10$). The 9 single-line ($H=1$) observations have mean $|z|$ of $1.38$,
within the range of the concentration deciles' means ($0.77$ to $1.52$); with 9 observations that
is a weak reading. A null diagnostic is consistent with the exponent absorbing the size
heteroscedasticity, but it cannot establish that it does. Rescale-invariant shape comparisons
(Anderson--Darling across groups, with Bowley-skewness, tail-skew and Moors-kurtosis comparisons)
do not establish drift along either axis. The row-level Anderson--Darling reference is not
calibrated for panel dependence or estimated group standardisation; the cluster-bootstrap
comparisons are descriptive.
Different individual scales within a group can change the shape of its mixture even when the
conditional shock shape is common, so these grouped comparisons do not isolate conditional
shape drift. They do not establish shape invariance or test the zero-location restriction (\supp{16}).

The clean values of $z$ lean adverse in the far tail: of the 605, 24 lie above $+3$ and 3 below
$-3$. These are descriptive counts. A Student-$t$ reference with the clean tail index ($\nu=3.96$)
would put about 12 in each tail, but $z$ omits the year and RITC scale terms, so that reference is
not the model's exact expectation. The 27 extremes come from 19 syndicates (Syndicate 1991 has 6 of
them) and cluster by reporting year, so no sign test that treats them as independent is reported;
the lean in the body of the distribution is not resolved (\supp{20}). The skew-$t$ refit of
Section~\ref{sec:results:posterior} moves the stress little because $605$ of the $632$ donors are
clean, and a clean donor's transferred value does not use the fitted shape of the shock. So this
check does not test how the adverse lean would act on the tail of the clean pool.

Held-out central plug-in prediction intervals are close to nominal (Table~\ref{tab:ooscal}). Central
coverage is a different question from one-sided coverage of the extreme tail, which the backtest
of Section~\ref{sec:results:backtest} addresses.

\begin{table}[H]
\begin{threeparttable}
\caption{Held-out calibration of central intervals: empirical coverage of the single-regime
composition model's central intervals, evaluated at each fold's posterior-mean scale and degrees of freedom and
pooled across the held-out folds.}
\label{tab:ooscal}
\begin{tabular}{lcc}
\toprule
\headrow Interval & Nominal & Held-out empirical \\
\midrule
90\% & 0.900 & 0.902 \\
95\% & 0.950 & 0.953 \\
99\% & 0.990 & 0.984 \\
\bottomrule
\end{tabular}
\end{threeparttable}
\end{table}

\subsection{Model choices and the evidence for each}
\label{app:robust:summary}
Table~\ref{tab:rejected} lists the model choices, the test of each and its result. Its status column has four
words: \emph{fitted} (the alternative was fitted and scored or estimated), \emph{diagnostic} (the adopted model was
checked without an alternative), \emph{structural} (a modelling decision the data do not test) and \emph{untested}. Among the fitted comparisons, five alternatives are not distinguished
from the adopted choices: no floor, $k=\tfrac12$, the effective-size concentration form, and the
long-tail share and dominant line added to the dispersion model.

\begin{table}[!b]
\begin{threeparttable}
\caption{Model choices, empirical comparisons and untested extensions.}
\label{tab:rejected}
\footnotesize
\renewcommand{\arraystretch}{0.95}
\begin{tabular}{>{\raggedright\arraybackslash}p{0.23\linewidth}>{\raggedright\arraybackslash}p{0.10\linewidth}>{\raggedright\arraybackslash}p{0.57\linewidth}}
\toprule
\headrow Alternative & Status & Result \\
\midrule
Sequential two-stage fit (size, then concentration) & structural & Ordering bias and circularity; one joint fit is used \\
Gaussian (least-squares) likelihood & structural & Tails are heavy ($\nu_{\mathrm{clean}}\approx3.96$); a Student-$t$ likelihood is used \\
One shock shape at every size and concentration & diagnostic & Grouped shape comparisons do not isolate conditional drift or establish invariance (\ref{app:robust:fit}) \\
No floor (pure power law) & fitted & Not separated: +1.35 [\textminus2.6, +5.5], $P=0.75$; larger exponent, $k=0.745$. The floor is retained as a prudential extrapolation: the floorless law tends to zero volatility as a fraction of reserves beyond the largest observed book, \pounds4{,}750m (\ref{app:robust:floor}) \\
Constant scale ($k=1$) & fitted & Strongly disfavoured: it is the naive market pool; +33.3 [+9, +63], $P=0.9972$; the size-only model leads it by +34.4 [+11, +64] (\ref{app:robust:pooling}) \\
$k=\tfrac12$ (finite-variance benchmark for the diversifiable term) & fitted & Not separated: +1.24 [\textminus4.0, +6.2], $P=0.70$; fixing $k$ there moves the worked stress from 0.342 to 0.315 (\ref{app:robust:pooling}) \\
Concentration overlay against size-only & fitted & Not separated: \textminus0.96 [\textminus3.8, +1.8], $P=0.24$; the default transfer is size-only (Section~\ref{sec:results:oos}, \supp{19}) \\
Separate power of $H$ ($H^{\delta}$) & untested & No such model was fitted; the data say nothing either way \\
Long-tail share in the scale & fitted & Not separated: the slope's 95\% HDI is above zero (+0.56 [+0.25, +0.87] in the adopted model), but the held-out gain of adding it is unresolved (+0.6 [\textminus11.0, +10.3], $P=0.57$); not carried by the transfer (\supp{19}) \\
Dominant line in the scale & fitted & No resolved effect: the between-line scale's HDI reaches zero and every line multiplier includes one (\supp{19}) \\
RITC treatment (exclusion, unadjusted, rank map) & fitted & $k$ and the floor stable across four treatments; no RITC donor sits at the worked target's 99.5\% point, so the rank map is not tested there (\supp{15}) \\
Skewed shock (skew-$t$) & fitted & $\delta=0.124$ [0.088, 0.162]; moves the worked stress by \textminus1.1\%, inside the 5\% decision line, so the shock stays symmetric (\supp{20}) \\
A location for the target & structural & Not estimated: donor movements carry any persistent level, scaled if clean and rank-mapped if RITC (\supp{16}) \\
Concentration-dependent location ($\mu=m_0+m_1H$) & fitted & Pooled slope negative (\textminus0.042, 95\% HDI [\textminus0.071, \textminus0.009]) but unresolved once syndicate structure is allowed; the within-book slope, the one a transfer would use, is not resolved: +0.006, 95\% HDI [\textminus0.049, +0.061]; not carried by the transfer (\supp{16}) \\
Chronological held-out split & untested & The out-of-sample checks hold out syndicates, not years (Section~\ref{sec:implications}) \\
\bottomrule
\end{tabular}
\begin{tablenotes}[hang]
\item[] Each score difference is held-out log predictive density on the five syndicate folds, from single-regime
fits that carry the concentration term with $\gamma$ estimated: the adopted form's score minus the alternative's
(for the long-tail share, the model with the share minus the model without; for the concentration row, the
effective-size form minus size-only). Intervals on a score difference are $95\%$ credible intervals and $P$ the
posterior probability that the first-named model predicts better, both under the Bayesian bootstrap over syndicates.
Intervals on a parameter (the long-tail slope, $\delta$ and the location slopes) are $95\%$ posterior HDIs.
\end{tablenotes}
\end{threeparttable}
\end{table}

\end{document}